\documentclass{jpp}
\usepackage{graphicx,bm,url,xcolor,placeins}
\graphicspath{{./fig/}{./png/}}
\newcommand{\EQ}{\begin{equation}}
\newcommand{\EN}{\end{equation}}
\newcommand{\EQA}{\begin{eqnarray}}
\newcommand{\ENA}{\end{eqnarray}}
\newcommand{\eq}[1]{(\ref{#1})}

\newcommand{\Eq}[1]{(\ref{#1})}

\newcommand{\Sec}[1]{\S~\ref{#1}}

\newcommand{\Fig}[1]{Figure~\ref{#1}}

\newcommand{\Figs}[2]{Figures~\ref{#1} and \ref{#2}}
\newcommand{\Figss}[2]{Figures~\ref{#1}--\ref{#2}}
\newcommand{\Tab}[1]{Table~\ref{#1}}
\newcommand{\Tabs}[2]{Tables~\ref{#1} and \ref{#2}}
\newcommand{\Tabss}[2]{Tables~\ref{#1}--\ref{#2}}

\newcommand{\bra}[1]{\langle #1\rangle}

\newcommand{\mean}[1]{\overline #1}

{}
{}

{}
{}

{}
{}
{}
{}
{}
{}
{}
{}
{}
{}
{}
{}
{}
{}

{}

\newcommand{\meanB}{\overline{B}}
\newcommand{\meanBrms}{\overline{B}_\mathrm{rms}}
\newcommand{\meanJrms}{\overline{J}_\mathrm{rms}}

\newcommand{\meanJ}{\overline{J}}

{}

{}
{}
{}

\newcommand{\zzz}{\hat{\mbox{\boldmath $z$}} {}}

\newcommand{\meanBB}{{\overline{\bm{B}}}}
\newcommand{\meanJJ}{{\overline{\bm{J}}}}

\newcommand{\BB}{\bm{B}}

\newcommand{\JJ}{\bm{J}}

\newcommand{\AAA}{\bm{A}}

\newcommand{\uu}{\bm{u}}

\newcommand{\nab}{{\bm{\nabla}}}

\newcommand{\ii}{{\rm i}}

\newcommand{\dd}{{\rm d} {}}

\def\Pm{\mbox{\rm Pr}_{\rm M}}

\def\kf{k_{\rm f}}

\def\Brms{B_{\rm rms}}

\def\Jrms{J_{\rm rms}}

\def\urms{u_{\rm rms}}

\newcommand{\Gyr}{\,{\rm Gyr}}

\title{Convergence of Roberts flow dynamos in Eulerian and Lagrangian codes}

\author[Shchutskyi, Brandenburg \& Schaller]{
Nikyta Shchutskyi\aff{1,2}\corresp{\email{shchutskyi@lorentz.leidenuniv.nl}},
Axel Brandenburg\aff{3,4,5,6}, \and\, 
Matthieu Schaller\aff{1,2}
}
\affiliation{
\aff{1}Lorentz Institute for Theoretical Physics, Leiden University, PO Box 9506, NL-2300 RA Leiden, The Netherlands
\aff{2}Leiden Observatory, Leiden University, PO Box 9513, NL-2300 RA Leiden, The Netherlands
\aff{3}Nordita, KTH Royal Institute of Technology and Stockholm University,
Hannes Alfv\'ens v\"ag 12, SE-10691 Stockholm, Sweden
\aff{4}The Oskar Klein Centre, Department of Astronomy,
Stockholm University, AlbaNova, SE-10691 Stockholm, Sweden
\aff{5}McWilliams Center for Cosmology \& Department of Physics,
Carnegie Mellon University, Pittsburgh, PA 15213, USA
\aff{6}School of Natural Sciences and Medicine, Ilia State University,
3-5 Cholokashvili Avenue, 0194 Tbilisi, Georgia
}

\date{\today}

\begin{document}

\maketitle
 
\begin{abstract}
Standard benchmark tests for numerical magnetohydrodynamics codes focus on one- and
two-dimensional problems that ignore the possibility of dynamos,
i.e., an exponential instability converting kinetic energy into magnetic energy.
They also tend to ignore magnetic helicity conservation that can have
dramatic effects in periodic domains.
The existence of magnetic diffusion is crucial for dynamos, as was
exemplified by the impossibility of finding dynamo solutions when using
Euler potentials, which is a perfectly valid approach in the absence of
magnetic diffusion.
The Roberts flows~I, II, III, and IV are two-dimensional flows that support
three-dimensional dynamos, all of which are also large-scale dynamos
in the sense that their planar averages are of significant strengths,
even though flow~II is pointwise non-helical.
Here we demonstrate the numerical convergence properties of solutions
obtained using discretisation schemes of second, sixth, and tenth order in
the mesh width in the \textsc{Pencil Code}, compare the results with the SPH-based \textsc{SWIFT} code, and present a list of averaged quantities that facilitate
their use in characterising the solutions as benchmarks. In addition, we present a procedure to obtain two-dimensional time-independent magnetic field visualisation for all Roberts flows, and perform per-pixel comparison between the \textsc{Pencil} and \textsc{SWIFT} codes.
The rich combination of features makes Roberts flows an excellent dynamo benchmark
for the direct numerical simulations of magnetohydrodynamics equations. 

\end{abstract}

\section{Introduction}

Magnetic fields are present in a wide range of systems, varying in size and magnetic field strength, and are important for astrophysics and cosmology.
The field is amplified by orders of magnitude inside structures such as stars, planets, and galaxies,
where the magnetic fields are compressed and then strongly amplified by the dynamo instability.
Dynamos efficiently convert the kinetic energy of the plasma into magnetic fields until the backreaction from the Lorentz force becomes significant.
The amplification happens due to turbulent eddies inside the plasma, 
creating a \textit{small-scale dynamo} that produces magnetic fields roughly on the scale of the turbulent eddies.  Interactions of such eddies can drive magnetic fields to much greater scales, supporting the \textit{large-scale dynamo}. 
This process can cover many orders of magnitude in spatial scales.
Because of such complexity, numerical codes are routinely used to obtain solutions to the underlying equations.
Since the magnetic and fluid Reynolds numbers are typically extremely
high, from $10^3$ up to $10^{17}$ \citep{Schober_2012}, one often neglects dissipative effects and works with what is
referred to as idealised magnetohydrodynamics (MHD). 
Formally, the absence of resistivity in ideal MHD removes the small-scale cutoff in the magnetic field spectrum.
Therefore, initially smooth magnetic fields can develop arbitrarily small features with time.
Then numerical discretisation effects can strongly impact the magnetic field evolution, for example, in the best case,
by introducing an effective numerical diffusion that results in deviations from ideal MHD, or in the worst case, a false dynamo, e.g., due to a numerical instability.

In the special case of a steady flow, \cite{MP85} have proved that the
growth rate of a dynamo must vanish when there is no magnetic diffusivity.
Thus, by modelling dissipative effects solely as a numerical device,
certain properties of the dynamo solution may have been compromised.
A very striking example is the use of Euler potentials, which allow one
to obtain correct solutions to the ideal MHD equations, but not to the
non-ideal ones \citep{Bra10}.
Dynamos appear therefore impossible to solve with Euler potentials.

An important question concerns the accuracy of various terms entering
the numerical solutions of the dynamo problem.
This can be critical when the magnetic Prandtl number $\Pm$ (the ratio of
kinematic viscosity to magnetic diffusivity) is very different from unity.
A specific problem related to a large value of $\Pm$ is the possibility
of a slowdown of the timescale of decaying MHD turbulence in
the early Universe during the radiation-dominated era \citep{HS23}.
Contrary to theoretical expectations, an extreme dependence on the
value of $\Pm$ has not been observed in the corresponding simulations
of \cite{Bra+24}, but there is also the possibility that this is due to
inaccuracies resulting from insufficient numerical resolution.
Meanwhile, earlier results of a $\Pm$ dependence of the reconnection rate in driven reconnection \citep{Comisso+15}
have now been challenged based on results in spontaneous reconnection \citep{KB26}.

Another prominent example is numerical simulations of isolated Milky Way-size galaxies.
On the one hand, dynamos should grow and saturate within galaxies to match observed field strengths.
While some simulations show fast resolution-dependent growth on timescales of $0.1$--$1\Gyr$,
followed by saturation even without supernova feedback \citep{Pakmor_2013, Wissing_2023}, others do not yield saturation,
but rather show shear-like linear growth, and only with the supernova feedback switched on do they observe a resolution-dependent exponential growth and saturate \citep{Rieder_2016}.
This is a picture that is more consistent with theoretical expectations \citep{Ruzmaikin1988}.
Studying such a regime is especially relevant in the context of galactic dynamos, because it could reveal spurious dynamo effects \citep{Ugalino_2026}.
On the other hand, studies of turbulent boxes in a regime with numerically induced resistivity indicate similar behaviour for different numerical implementations \citep{Pakmor_2026}. Generally, due to the very broad range of scales involved in isolated galaxy simulations, they require additional sub-grid turbulence models \citep{Vazza_2020,Liu_2022}. Such disagreement indicates the existence of uncertainty in dynamo action due to the discretisation method used for isolated galaxy simulations.

An additional source of error that is commonly discussed in the literature
\citep{HH24} is the possibility of spurious divergences in the magnetic field, $\bm{B}$.
Their effect on the dynamo is hard to assess in general.
However, this source of error can be eliminated by evolving the magnetic vector potential \citep[e.g.][]{BRANDENBURG2002471,STASYSZYN2015148},
or it can be reduced through divergence cleaning \citep[e.g.][]{POWELL1999284,DEDNER2002645}.
Another way is designing the discretisation specifically to conserve $\mathrm{div}\bm{B}$ by employing the constrained transport method \citep{Evans_1988}.

To quantify the accuracy of a particular dynamo solution, one can compare
characteristic quantities with a reliable reference solution.
Examples include the growth rate, the resulting mean magnetic
field, the magnetic Taylor microscale, or the rate of work done against
the Lorentz force, which provides the energy gain of the dynamo against
dissipation.

In the present paper, we use the four Roberts flows \citep{Rob72}, which represent a set of simple 2D systems, well studied semi-analytically and with direct numerical simulations, each of them illustrating a specific regime of a large-scale dynamo growth. Roberts flows are what is called \textit{slow dynamos},
i.e., their growth rates go to zero in the limit of large magnetic Reynolds numbers.
We employ two completely different codes, \textsc{Pencil}, a grid-based Eulerian code which has been
frequently used for various dynamo problems \citep{JOSS}, and \textsc{SWIFT} \citep[][]{SWIFT},
a Lagrangian particle-based cosmological code for which the performance was tested in one of the four Roberts flows (flow~I) and in an ABC flow setup \citep{SSKSB25}.
While \textsc{Pencil} evolves the magnetic fields using the vector potential with a suitable gauge at fixed grid points,
\textsc{SWIFT} evolves the magnetic field directly, employs divergence cleaning to ensure a physical evolution of the magnetic field,
and solves the equations at particle positions in a frame moving with the fluid.
We cross-check the results using both codes and evaluate possible differences caused by the choice of discretisation.
We investigate the dependence of the growth rate on resistivity (or magnetic Reynolds number), and study the convergence of the quantities of interest with resolution for the Roberts flow I at a specifically chosen resistivity.
To compare the results, we address the question of a unique visualisation of the growing 3D magnetic fields for all four flows.
With this data, Roberts flows can be used as benchmarks for code accuracy in reproducing dynamos in the kinematic regime.
They also provide an additional tool for performing astrophysical and cosmological MHD code comparison.

\section{Problem definition, excitation conditions, and numerical methods}
\subsection{The four Roberts flows}

\cite{Rob72} proposed four 2D dynamo capable flow patterns.
They are kinematic flows, i.e., the Navier-Stokes equation is not solved.
The velocity depends only on $x$ and $y$, but all three components of
the velocity vector $\bm{v}$ are non-vanishing.
Flow~I, for example, can be written as $\bm{v}=\kf\varphi\zzz+\nab\times(\varphi\zzz)$.
More explicitly, the four flows have the same velocity in the $ x$ and $y$ directions and have different velocities in the $z$ direction.
\begin{equation}
\bm{v}(x,y)=\pmatrix{
+v_0\sin k_0x \cos k_0y\cr
-v_0\cos k_0x \sin  k_0y \cr
v_z (x,y)
}.
\end{equation}
The choice determines the dynamo growth mechanism, excitation conditions, and solution properties (\Tab{Tflows}).
The parameters $v_0$ and $k_0$ are constants determining the flow velocity and wavenumber,
and the choice of $v_z(x,y)$ determines the magnitude of the velocity in the $z$ direction relative to the in-plane velocity.
The form and amplitude of the characteristic $z$ velocity, $w_0 = v_0 \kf/k_0 $, will be varied.
As discussed elsewhere \citep{Tilgner_2008,Delven_2013,Rhei+14}, the growth can be studied in terms of the $xy$ planar averaged magnetic field:
\begin{equation}
\meanBB(z,t)=\int\BB(x,y,z,t)\,\dd x\,\dd y/L_\perp^2.
\end{equation}
The electromotive force for a planar averaged magnetic field can be described as \citep[e.g.,][]{BRS08}
\begin{equation}
\mean\mathcal{E}_{i}(z,t) = \int d \xi \int_{-\infty}^t d\tau \left[ \hat{\alpha}_{ij}( \xi,\tau) \meanB_{j}(z-\xi,t-\tau) - \hat{\eta}_{ij}( \xi,\tau)\meanJ_{j}(z-\xi,t-\tau) \right],
\end{equation}
where $\hat{\alpha}_{ij}$ and $\hat{\eta}_{ij}$ are integral kernels
describing the total effective response of the electromotive force to the mean planar averaged magnetic field and magnetic current correspondingly.
The integral form allows for spatio-temporal non-locality, i.e., the response depends on the neighbouring points and past times \citep{HB09}.
The local forms correspond to $\delta$ functions, i.e., $\hat{\alpha}_{ij}=\alpha_{ij}\, \delta(\xi)\delta(t)$ and $\hat{\eta}_{ij}=\eta_{ij}\delta(\xi)\delta(t)$.
However, the nonlocal forms are crucial for understanding the dynamo effects for Roberts flows~II--IV.
Applying a Fourier transform, we get the electromotive force as
\begin{equation}
\mean\mathcal{E}_{i}(\omega, k_z) = \tilde{\alpha}_{ij}(\omega, k_z) \mean B_{i}(\omega, k_z) - \tilde{\eta}_{ij}(\omega, k_z) \mean J_{i}(\omega, k_z). 
\end{equation}
Thus, the Fourier-transformed kernels $\tilde{\alpha}_{ij}(\omega, k_z)$ and $\tilde{\eta}_{ij}(\omega, k_z)$ have now become multiplicative coefficients
that depend on frequency $\omega$ and wavenumber $k_z$.
The choice of the velocity in the $z$ direction impacts the electromotive force by changing the form of the $\hat{\alpha}_{ij}$ and $\hat{\eta}_{ij}$ kernels.
With the \textit{alpha effect}, the magnetic field amplification proceeds
due to the electromotive force being proportional to the mean magnetic field, $\mean\mathcal{E}_i \sim \tilde{\alpha}_{ij} \mean B_j$.
The $\tilde{\alpha}_{ij}$ tensor can have different particular forms:
it can be diagonal, $\tilde{\alpha}_{ij} = \alpha \,\delta_{ij}$; off-diagonal symmetric, $\tilde{\alpha}_{12}=\tilde{\alpha}_{21} = c $, $\tilde{\alpha}_{11}=\tilde{\alpha}_{22} = 0 $,
or it could be antisymmetric of the form $\tilde{\alpha}_{12}=-\tilde{\alpha}_{21} = \gamma $, $\tilde{\alpha}_{11}=\tilde{\alpha}_{22} = 0 $.
In the latter case, the electromotive force takes the form
$\boldsymbol{\mean \mathcal{E}} = ...+\gamma \hat{\boldsymbol{e}}_z\times  \boldsymbol{\meanB}$,
which is equivalent to an additional turbulent velocity contribution, also referred to as \textit{turbulent pumping}.
The electromotive force also depends on the total resistivity tensor $\mean\mathcal{E}_i =... -\tilde{\eta}_{ij} \mean J_j$.
For flows I--IV, the resistivity tensor stays diagonal, i.e., $\tilde{\eta}_{ij}= \eta_\mathrm{tot} \, \delta_{ij}$.
The total resistivity has two contributions: Ohmic diffusion and effective turbulent diffusivity $\eta_\mathrm{tot} = \eta_{\rm ohm} +\eta_\mathrm{turb}$.
Under some conditions, the turbulent part can become negative on large length scales, but remains positive for small-scale magnetic fields.
When the negative turbulent contribution is greater than the Ohmic dissipation, a mean-field dynamo from \textit{negative turbulent diffusivity} occurs \citep{Delven_2013}.

When both tensors have frequency-dependent complex contributions, for example, $\tilde{\alpha}_{ij},\tilde{\eta}_{ij} \sim (1 + i\omega\tau)^{-1}$,
where $\tau$ is a relaxation time that is related to the turnover time,
the electromotive force depends on the magnetic field's history, leading to the \textit{memory effect } \citep{HB09}.
In special cases, this can lead to dynamo action \citep{Rhei+14}.
The four Roberts flows explore these regimes of the mean-field slow dynamo.
 
The conditions to excite growing solutions for the four flows vary significantly with the domain geometry, mode wavenumber, and magnetic Reynolds number.
Under some conditions, dynamos cannot be excited.
For example, in a cubic domain for flows~II and III, when $k_0$ is too small, e.g., $k_0=1$ (that is, when the $xy$ plane contains 4 vortices).
It is then necessary to elongate the $z$ direction such that the mode $k_z < 0.64 \, k_0$ (for flow II) or $k_z < 0.78 \, k_0$ (for flow III) can be excited.
For this reason, it is convenient to choose an elongated domain with $L_\parallel = 2 L_\perp$, corresponding to $k_z=0.5 \, k_0$,
where all four flows can be excited \citep{Rhei+14,Ntormousi_2023}.
The approximate onset Reynolds numbers for such a setup are shown in \Tab{Tflows} \footnote{A possible explanation for the $R_M^{\rm crit}$ discrepancy is an unaccounted-for $v_{\rm rms}$ factor in the calculation of $R_M^{\rm crit}$ for flows~I--III. For flow~IV, where $v_{\rm rms}=1$, our values of $R_M^{\rm crit}$ agree with those reported in the reference. However, we were unable to verify this explanation because the reference does not provide the corresponding values of $\eta_{\rm crit}$.}.

Flows I and IV have a steady normalised magnetic field configuration, with the plane-averaged magnetic field being coupled in the case of flow I and independent for flow IV.
For flow II (III), the solutions move in the opposite (same) $z$ directions.

The growth rates for the modes also depend on the choice of the constant in front of the velocity term in the $z$ direction.
Typical choices are $w_0=v_0$ and $w_0 = v_0 \sqrt{2}$.
Both choices are explored for flows~I and II, but the value $w_0=v_0$ was used to compare the four flows.
The choice $w_0=v_0$ leads to $v_{\rm rms} = \sqrt{3}/2 \simeq 0.866$ for flows~I, II, and III,
and $v_{\rm rms} = 1$ for flow~IV, while choosing $w_0=\sqrt{2}v_0$ leads to $v_{\rm rms} = 1$ for flows~I and II.

\begin{table}\caption{
Summary of the dynamo mechanisms in Roberts flows.
The table shows the form of the velocity in the $z$ direction, the form of the $\tilde{\alpha}_{ij}$ tensor, the mechanism driving the growth,
features of spacial magnetic field distribution $B_i(\bm  r, t)/B_\mathrm{rms}(t)$ for the growing modes, coupling of the in-plane mean field evolution $\mean B_x\,\&\,\mean B_y$,
and the critical magnetic Reynolds number needed to excite a dynamo for the elongated domain ($L_\parallel = 2L_\perp$), with the choice $w_0=v_0$ of velocity in $z$ direction.
This is based on a similar table in \cite{Ntormousi_2023}, where for flows I, II, and III the values of $R_{\rm M}^\mathrm{crit}$ were larger
because the factor of $\sqrt{3}/2 \simeq 0.866 = v_{\rm rms}$ was not taken into account.
}
\vspace{12pt}\centerline{\begin{tabular}{c|c|c|c|c|c|c}

Flow &
$v_z(x,y)/w_0$ &
$\tilde{\alpha}_{ij}(\omega,k)$ &
Mechanism & $\BB_0(x,y,z,t)$ & $\mean B_x \,\&\,\mean B_y$
& $R_{\rm M}^\mathrm{crit}$
\\

\hline 
$\rm I$ &
$\sin k_0x\,\sin k_0y$ &
$
\pmatrix{
\alpha & 0\\
0 & \alpha
}$ &
$\alpha$ effect
 & Steady
 & Coupled
 & $1.75$
\\[1em]

$\rm II$  &
$\cos k_0x\,\cos k_0y$ &
$\pmatrix{
0 & a\\
a & 0
}
$
&
\begin{tabular}{@{}c@{}}

Off-diag. $\tilde{\alpha}_{ij}$ \\
+ memory eff.
\end{tabular}
& 
\begin{tabular}{@{}c@{}}
2 opposite waves\\ in $z$ direction 
\end{tabular}
& Indep.
& $5.97$

\\[1em]

$\rm III$  
&
$\frac{1}{2}
(\cos 2k_0x+\cos 2k_0y)$ 
&
$
\pmatrix{
0 & \gamma\\
-\gamma & 0
}$
&
\begin{tabular}{@{}c@{}}
Turb. pump. \\
+ memory eff. \\
\end{tabular}
& \begin{tabular}{@{}c@{}}
2 parallel waves \\ in $z$ direction
\end{tabular}
& Indep.
& $3.40$
\\[1em]

$\rm IV$  &
$\sin k_0x$ &
$ 0 $
&
\begin{tabular}{@{}c@{}}
Negative turb.\\
diff., $\eta_\mathrm{turb}<0$ 
\end{tabular}
& Steady 
& Indep.
& $4.54$
\\[1em]
\label{Tflows}\end{tabular}}\end{table}

\subsection{Eigenfunctions of the solution} \label{sec:growing_solutions}
We are looking for growing solutions to the induction equation,
\begin{equation}
\frac{\partial\bm{B}}{\partial t}=\bm{\nabla}\times(\bm{v}\times\bm{B})
+\eta\nabla^2\bm{B},
\label{InductionEquation}
\end{equation}
where $\eta$ is the magnetic diffusivity.
For a growing solution, $\BB(x,y,z,t)$ is of the form
$\BB_0(x,y,z)\,e^{\lambda t}$, where $\lambda$ is the eigenvalue and $\BB_0(x,y,z)$ is the eigenfunction.
Note that for flows II and III, $\lambda$ is complex and the solutions correspond to waves travelling in the positive or negative $z$ direction.
In those cases, $\BB_0(x,y,z)$ is also complex.
To compensate for the growth, the magnetic field can be normalised by the root-mean-square value over the box:
\begin{equation}
  \bm{B}_0(x,y,z,t) = \bm{B}(x,y,z,t)/B_{\rm rms}(t).
\end{equation}
We note that there is a residual time dependence for flows II and III because the solutions are travelling waves.
Since the velocity does not depend on $z$, the solution can be written as
$\BB_0(x,y,z,t)=\BB_{0\,\perp}(x,y,t)\,e^{\ii k_z z}$ with a complex function
$\BB_{0\,\perp}(x,y,t)$.
It can be expressed as:
\begin{equation}
\tilde{B}_{0i}(x,y,t)=r_i(x,y,t)\,e^{\ii\theta_i(x,y,t)}, \quad i=1,2,3,
\label{ModulusPhase}
\end{equation}
where $r_i(x,y,t)$ and $\theta_i(x,y,t)$ are the moduli and phases of the
(two-dimensional) eigenfunction.
The phase ambiguity is fixed for the plane-averaged field in the $x$ direction:
\begin{equation} \tilde B_{0i}(x,y,t) \rightarrow e^{i\phi} \tilde B_{0i}(x,y,t), \qquad
   {\rm Im} \left( e^{i\phi}  \tilde{\overline{B}}_{0x}\right) = 0 \quad \rightarrow \quad \theta_i(x,y).
\end{equation}
This choice also eliminates contributions to the time dependence of the phase from travelling waves of the form $\phi=\omega t$

\subsection{Single mode separation for Roberts flow II and III} \label{sec:time_independent_patterns}
The growing solutions for the Roberts flows II and III contain two modes travelling in the $z$ direction \citep{Rob72, Rhei+14}. The waves have two polarisations:
\begin{equation}
    \overline{\bm{B}_1}(z,t) = \bm{e}_x B_{x,0}\sin(k_z z - \omega t), \, \, \,
     \overline{\bm{B}_2} (z,t) = \bm{e}_y B_{y,0} \sin\left(k_z z - (-1)^n\omega t\right)
\end{equation}
where $n=1$ for Roberts flow II (waves travel in opposite directions in $z$) and $n=0$ for Roberts flow III (waves propagate in the same direction). 
Separating the modes is useful because the magnetic field for both modes locally depends on $x,y$ and time, leading to interference. To illustrate this, the local magnetic field for the two modes can be written as:
\begin{equation}
    \bm{B}(x,y,z,t) = e^{\lambda t}\left(\bm{b}_1(x,y) e^{i\omega t} + \bm{b}_2(x,y) e^{i(-1)^n\omega t}\right) e^{ikz} + \mathrm{c.c.}
\end{equation}
where $\mean b_{1,y} = 0$, $\mean b_{2,x} = 0$.
Following the same procedure as in \Eq{ModulusPhase}, the space- and time-dependent amplitude can be expressed as:
\begin{equation}
    r_i(x,y,t) = \left|r_{1,i}(x,y)  + r_{2,i}(x,y) e^{i((-1)^n-1)\omega t+i(\theta_{2,i}-\theta_{1,i})(x,y)}\right |,
\end{equation}
where we decomposed each mode according to \Eq{ModulusPhase}.
While for flow~III ($n=0$) there is no time dependence for $r_i(x,y,t)$, there is a time-dependent interference for flow~II ($n=1$).
However, the choice of initial condition, such as the sine wave in the $z$ direction for a perpendicular field,
similar to those used by \cite{Rob72}, can be used to remove the interference entirely for the Roberts flows~II and~III:
\begin{equation}
    \bm{B}_0(x,y,z) = \bm{e}_x B_{0} \sin(k_0 z),
    \label{equation:initial_condition}
\end{equation}
where $\bm{b}_2(x,y)=0$, since $\mean B_y = 0$.
These initial conditions were used only for the time-independent magnetic field pattern studies.
(In all other studies, a random vector potential field was used as initial condition.)
In the following, the time-independent patterns are obtained in both codes for $r_i$ and $\theta_i$.

\subsection{Observables for the growing solutions} \label{sec:observable_characteristic_quantities}

We determine the growth rate by computing a linear fit of $\ln B_\mathrm{\rm rms}(t)$ as a function of time
at a sufficiently late time when the exponential growth has commenced,
but the numerical values have not yet exceeded the values above which
the numerical precision deteriorates. For $B_{\rm rms}$, which oscillates with time, we follow \cite{SSKSB25}: we locate peaks in the local growth rate, $d/dt\ln B_\mathrm{\rm rms}(t) $ and fit the peaks to extract the growth rate $\lambda$.
Frequencies are measured by counting the separation between the peaks, $T$; the frequency is $\omega_{\rm mod} = 2\pi/T_{\rm mod}$.
(The oscillation frequency of individual magnetic field components half as big, i.e.\ $\omega = \omega_{\rm mod}/2$.)
In addition to $\lambda$,
we also determine quantities that characterise the eigenfunction.
Of particular interest is the rms value of the $xy$-averaged mean
field, $\meanBrms=\bra{\meanBB^2}^{1/2}$ illustrating the magnitude of the in-plane large scale field, $\bar B_{\rm rms}/B_\mathrm{rms}$.
Here, angle brackets denote volume averaging.
This quantity can be used to find the relative contribution of the in-plane large-scale field to the volume-averaged rms magnetic field.
Other quantities of interest are the magnetic
Taylor microscale wavenumber, $\Jrms/\Brms$, the mean current
helicity, $\bra{\JJ\cdot\BB}$, as well as the rms value of the current
helicity, $\bra{(\JJ\cdot\BB)^2}^{1/2}$, the work done against
the Lorentz force $W_\mathrm{L}=\bra{\uu\cdot(\JJ\times\BB)}$, and
the length scale, $\Jrms/\Brms$. These quantities are exponentially
increasing with time.  It is therefore useful to normalise them by $\Brms$
or $\Brms^2$, depending on whether the quantities are linear or quadratic
in the magnetic field.

It might seem surprising that $\bra{\JJ\cdot\BB}=0$ for the marginally excited state.
This is true even for Roberts flow~I, whose kinetic helicity is maximal.
However, the vanishing of $\bra{\JJ\cdot\BB}$ is simply because of
magnetic helicity conservation in a periodic domain,
\begin{equation}
\dd\bra{\AAA\cdot\BB}/\dd t=-2\eta\mu_0\bra{\JJ\cdot\BB}.
\end{equation}
It is important to note that for periodic boundary conditions,
$\bra{\AAA\cdot\BB}$ is gauge-invariant.
Therefore, $\bra{\AAA\cdot\BB}$ is a physically meaningful quantity,
whose time derivative must vanish in the statistically steady state.
This demands that also $\bra{\JJ\cdot\BB}=0$ in the steady state.

\subsection{\textsc{Pencil} and \textsc{SWIFT} codes}
In many astrophysical applications, the motion of an ionised conductive gas with magnetic fields is described with the magnetohydrodynamics (MHD) equations.
While the magnetic field evolution is described by \Eq{InductionEquation}, the evolution of the fluid follows \citep{BS05}:
\begin{equation}
    \frac{\partial\bm{v}}{\partial t} + (\bm{v} \cdot \nabla) \bm{v}  = -\frac{1}{\rho} \nabla P -\frac{1}{\rho}\nabla \boldsymbol{\mathsf{S}} +\bm{\Pi},
    \label{eq:FluidEquation}
\end{equation}
with $\bm{v}$ the fluid velocity, $\rho$ the fluid density, the vector $\bm{\Pi}$ corresponds to an additional physical or artificial viscosity, 
$\boldsymbol{\mathsf{S}}$ the Maxwell
stress tensor:
\begin{equation}
    S_{ij} =
    \frac{1}{2} \frac{B^2}{\mu_0} \delta_{ij}
    - \frac{B_{i} B_{j}}{\mu_0},
\end{equation}
with $\mu_0$ the vacuum permeability, and the thermal pressure:
\begin{equation}
  P = (\gamma - 1) \rho u
\end{equation}
where $u$ is the specific internal energy.
For kinematic dynamos, the magnetic field back-reaction is neglected, and the velocity field is imposed instead of being evolved; \Eq{eq:FluidEquation} is not solved,
and gas is kept at constant temperature and density.

Within the \textsc{Pencil} code \citep{JOSS}, the induction equation \eq{InductionEquation} is solved by evolving the vector potential $\bm{A}$:
\begin{equation}
    \partial_t 
    \bm{A} = \bm{v} \times {\rm curl} \bm{A} + \eta \Delta \bm{A} - \nabla \Gamma,
\end{equation}
where the Weyl gauge, $\Gamma=0$, is chosen.
The magnetic field $\bm{B}$ and the current density $\bm{J}$ are obtained by consecutive applications of curl operators:
\begin{equation}
    \bm{B} = {\rm curl} \bm{A}, \qquad \bm{J} = {\rm curl} \bm{B},
\end{equation}
where we have set the permeability to unity.
This choice excludes any influence of spurious numerical ${\rm div }\bm{B}$ on the magnetic field evolution.
The use of vector potentials eliminates the magnetic field divergence up to the precision of the chosen $\rm curl$ operator.
The equation is solved in a fixed frame of reference, \textit{Eulerian frame}.
The gradient operators are defined on a fixed uniform grid and calculated as a combination of values of a function at the point
and its neighbours in the direction of the derivative using a centred finite differences approach.
For example, for 6th-order accurate derivatives \citep{BRANDENBURG2002471}:
\begin{equation}
f_i' = (-f_{ i-3}+9f_{ i-2}-45f_{i-1}+45f_{ i+1}-9f_{ i+2}+f_{i+3})/(60 \delta x),
\label{Pencil_df}
\end{equation}
\begin{equation}
f_i'' = (2f_{ i-3}-27f_{ i-2}+270f_{ i-1}-490f_{ i}+
270f_{ i+1}-27f_{ i+2}+2f_{ i+3})/(180 \delta x^2).
\label{Pencil_d2f}
\end{equation}
The \textsc{Pencil} code also has 2nd-, 4th-, 8th- and 10th-order accurate spatial derivative implementations.
For the time integration, the third-order low storage $2N$ Runge--Kutta scheme is
used \citep{Williamson80}.\\ %\citep{Williamson80}.

In contrast, \textsc{SWIFT} uses the method of \cite{Karapiperis_2026}, built on top of the Smoothed Particle Hydrodynamics scheme of \cite{Borrow2022},
to evolve the density-weighted magnetic field at particle positions (shown here in continuous form):
\begin{equation}
    \frac{d}{dt} \frac{\bm{B}}{\rho} = \left(\frac{\bm{B}}{\rho}\cdot \nabla\right)\bm{v} + \frac{\eta}{\rho} \Delta \bm{B} - \frac{\nabla \psi}{\rho},
\end{equation}
or, equivalently, in terms of the magnetic field evolution:
\begin{equation}
    \frac{d}{dt} \bm{B} = \left(\bm{B}\cdot \nabla\right)\bm{v} - \bm{B} \,{\rm div} \bm{v}+\eta \Delta \bm{B} - \nabla \psi,
\end{equation}
where $\psi$ is the divergence cleaning field \citep{DEDNER2002645}.
This equation is obtained from \Eq{InductionEquation} by changing the frame of reference to the one moving with the fluid,
the \textit{Lagrangian frame}, and neglecting the $\bm{v}\, {\rm div} \bm{B}$ term, which vanishes in the continuum limit \citep{Price_2012}.
In this form, the induction equation is Galilean invariant, and the magnetic divergence is an advected quantity.
For dynamically unimportant magnetic fields, and in the absence of Dedner cleaning or physical resistivity, the magnetic field evolution at a given particle position does not depend on the magnetic field of neighbouring particles.
The cleaning field $\psi$ is sourced by the magnetic field divergence or matter compression.
It spreads the divergence with an appropriately chosen cleaning speed $c_{\rm h}$ and decays over the timescale $\tau$:
\begin{equation}
    \frac{d}{dt} 
    \psi  =
    - c_{\rm h}^2 {\rm div} \bm{B}
    - \frac{1}{2} \frac{\psi}{c_{\rm h}}{\rm div }\bm{v} 
    - \psi/\tau
    \label{eq:Dedner}.
\end{equation}
In the following Roberts flow dynamo runs with \textsc{SWIFT}, we disable the divergence cleaning to see the behaviour emerging from the interplay of induction and diffusion terms only.
\noindent Gradient operators are defined at particle positions using the Smoothed Particle Hydrodynamics (SPH) method \citep{SSKSB25, Karapiperis_2026}.
For example, the induction term at the position of particle $a$ is the weighted sum over neighbouring set of particles $b$
within a finite size bell-shaped kernel $W_{ab}$ with width $h_a$ and cutoff $\simeq 2h$:
\begin{equation}
\left(\bm{B}\cdot \nabla\right)\bm{v}\bigg|_{\bm{r}=\bm{r}_a} = -\frac{f_{ab}}{\hat{\rho}_{a}}\sum_{b} m_b(\bm{v}_{a}-\bm{v}_{b})(\bm{B}_{a} \cdot \nabla_{a}W_{ab} (h_{a})),
\end{equation}
where $f_{ab}$ is a kernel correction factor, $m_b$ and $\rho_b$ are the neighbouring particle's mass and density,
$v_a,v_b$ are the velocities of a particle and its neighbour, $B_a$ is the magnetic field at the particle location.  
The method is formally first-order accurate.
However, the true accuracy significantly depends on the particle arrangement \citep{Price_2012}.

The code stores the magnetic field, current, divergence, etc, at particle positions, but it does not compute the vector potentials.
Therefore, there is no direct way of computing $\bm{A} \cdot \bm{B}$.
For the time integration, the \textsc{SWIFT} code employs a first order accurate leapfrog scheme \citep{SWIFT}.

The two codes represent two extremely different discretisation of the induction equation.
One employs a fixed-grid Eulerian approach, while the other is a fully Lagrangian, particle-based method.
One is frame-dependent, while the other is Galilean invariant.
Moreover, one evolves the magnetic vector potential, while the other solves the induction equation directly with additional divergence cleaning.
By using these very different approaches to study kinematic dynamos,
we can better understand the systematic effects that different discretisations have
on the observed properties of kinematic dynamos in direct numerical simulations.

\subsection{Code comparison} \label{sec:code_comparison}
Because \textsc{swift} has no direct way of computing vector potentials, one alternative way to compute $\langle \bm{A} \cdot \bm{B}\rangle$ is to reconstruct $\bm{A}$ and also $\bm{J}$ from the magnetic field $\bm{B}$ in Fourier space. Because of periodicity, $\langle \bm{A} \cdot \bm{B}\rangle$ is gauge-independent, and any suitable gauge can be chosen for \textsc{swift}. This procedure can add systematic errors to the result, and therefore it is useful to obtain results for both $\textsc{pencil}$ and $\textsc{swift}$ by applying the same procedure and comparing with calculations using only built-in $\textsc{pencil}$ operators. Therefore, a grid map of the $\bm{B}$ field is needed for both codes.
 In \textsc{swift}, the particle data can be mapped onto a grid using the \textsc{swiftsimio} tool \citep{Borrow2020}. For example, the SPH smoothed magnetic field is defined as:\\
\begin{equation}
    \bm{B}(\bm{r}) = \sum_b \bm{B}_b W(\bm{r}-\bm{r}_b, h),
\end{equation}
where $\bm{r}$ is the position on a grid of interest, $\bm{B}_b$ is the local particle magnetic field, $\bm{r}_b$ is the neighbouring particle position,
$W$ is the bell-shape SPH kernel (the Wendland C2 kernel is used), and $h$ is the smoothing length.
The width of the kernel $h$ is such that the kernel keeps a fixed average number of neighbours $N_\mathrm{neigh}\sim 100$ within the kernel cutoff $d = \xi h$, within a $\xi\simeq 2$ sphere.

For \textsc{Pencil}, the data is already on a grid, but the magnetic field is stored as the vector potential, so the built-in $\rm curl$ operators were used to obtain the 3D magnetic field grid.
Once velocities and magnetic fields on the same-sized grid were defined for both codes, a discrete Fourier transform was applied to the magnetic field to obtain its Fourier image:
\begin{equation}
    \tilde{\bm{B}}(\bm{k}) = \sum_{\bm{r}} e^{-i \boldsymbol{k}\cdot \boldsymbol{r}} \,\bm{B}(\bm{r}),
\end{equation}
where the sum goes over each grid point $\bm{r}=(x_i,y_j,z_k)$.
The Fourier images for the vector potential and the current density can then be obtained as
\begin{equation}
    \tilde{\bm{A}(}\bm{k}) = \frac{ {\rm i} \bm{k}\times \tilde{\bm{B}}}{k^2}, \qquad \tilde{\bm{J}}(\bm{k}) = {\rm i}\bm{k}\times \tilde{\bm{B}}, \qquad |\bm{k}|\neq 0, \qquad \tilde{\bm{A}}(\bm{0})=\bm{0}.
\end{equation}
This way, the vector potential is defined in the Coulomb gauge, that is $\bm{k}\cdot \tilde{\bm{A}} = 0$.
This is different from the gauge choice used internally in the \textsc{Pencil} code.
However, this choice does not affect the results because the quantities of interest are gauge independent. 
Space dependent quantities $\bm{Q}(\bm{r})$ are obtained later by applying an inverse discrete Fourier transform:
\begin{equation}
    \bm{Q}(\bm{r}) = \sum_{\bm{k}} e^{i \boldsymbol{k}\cdot \boldsymbol{r}} \,\tilde{\bm{Q}}(\bm{k}).
\end{equation}
For the convergence study of a quantity of interest $X$ from \Sec{sec:observable_characteristic_quantities},
we denote the error as the modulus of the difference between the measured and the reference values, i.e.,
\begin{equation}
    \mathrm{err}(X)=\bra{|X-X_\mathrm{ref}|}_t
    \label{eq:error_def}
\end{equation}
for a quantity $X$
and its reference value $X_\mathrm{ref}$, where $\bra{\cdot}_t$ denotes
a time average over a time-compensated function, as defined in \Sec{sec:observable_characteristic_quantities}.
For the study of the convergence of the time-independent patterns with sinusoidal initial conditions,
we look at the absolute value of the complex magnetic field from \Eq{ModulusPhase}:
\begin{equation}
    |\tilde{\bm{B}}|^2 = r_1^2+r_2^2+r_3^2.
    \label{PatternAbs}
\end{equation}
The difference between two patterns can be quantified as the absolute value of the difference of two complex maps:
\begin{equation}
   | \delta \tilde{\bm{B}}| = | \tilde{\bm{B}}_{\rm SWIFT}  - \tilde{\bm{B}}_{\rm Pencil} |,
    \label{PatternDifference}
\end{equation}
where $\tilde B_{\rm SWIFT}$ is the magnetic field distribution for the \textsc{swift} runs,
and $ \tilde B_{\rm Pencil}$ is the magnetic field distribution for the \textsc{Pencil} 2nd order scheme.\\
The root-mean-square-averaged deviation over the $xy$ plane is used to quantify the overall pattern difference between the two codes:
\begin{equation}
    \langle |\delta \tilde{\bm{B}}|\rangle_{\rm rms}^2 = \frac{1}{N_\perp^2}\sum_{i,j} |\delta \tilde{\bm{B}}|^2(x_i,y_j).
    \label{PatternDifferenceRMS}
\end{equation}

\subsection{Spacial resolution limits for Roberts flows}\label{sec:spacial_resolution_limits}
The induction equation is solved in \textsc{SWIFT} by applying the SPH discretisation to the induction equation.
While the code can simulate, for example, the Roberts flow I dynamo, there is a limit on achievable magnetic Reynolds numbers at a given spatial resolution.
The main reason is that the induction term $(\bm{B}\cdot \nabla) \bm{v}$ generates smaller-scale features (due to winding), leading to the creation of a magnetic field cascade.
The resistive term $|\eta\Delta\bm{B}| \sim \eta k^2 B$ increases as the cascade reaches smaller scales (larger $k$) until the cascade stops.
There is no bound on the resistive source magnitude.
However, due to the discrete nature of gradients, exacerbated by the unstructured nature of the SPH method, the response is not linear but some function of the characteristic scale $k$, that is: 
\begin{equation}
    |\eta\Delta\bm{B}| \sim \eta k_{\rm eff}^2 B, \qquad k_{\rm eff} \simeq
    \cases{
    k, \qquad \quad  k \ll k_{\rm Ny} =  \pi/ d_\mathrm{res}, \cr
    c \cdot k_{\rm Ny}, \quad k\gtrsim k_{\rm Ny}.
  }
\end{equation}
where $k_{\rm Ny}$ is the Nyquist momentum scale, and $c\simeq 1$ is a constant prefactor.
When gradients approach the resolution scale, the effective wavevector stops growing, and the Laplacian operator
can no longer prevent the cascade from reaching the resolution scale and effectively turns off the resistivity.
The resulting magnetic field evolution is not smooth and can create spurious growth.
This is the overwinding regime.
In the case of Roberts flows, this can be expressed as a bound on the magnetic Reynolds number \citep[][]{SSKSB25}:
\begin{equation}
    R_{\rm M}  = \frac{v_{\rm rms}}{\eta k_0}\lesssim \frac{L_{\rm v}^2}{\pi d_\mathrm{res}^2} \simeq \frac{ k_{\rm Ny}^2}{\pi k_{\rm v}^2},
\end{equation}
where $L_{\rm v}$ is the single vortex scale, $d_\mathrm{res}$ is the effective resolution scale, and $k_{\rm v}=2\pi / (2 L_v)$.
For a $16^3$ resolution in Roberts flow runs, only $8$ particles resolve a single vortex, leading to constraints on the maximum permissable Reynolds number:
\begin{equation}
    R_{\rm M}^\mathrm{max}(N_\perp = 16)\sim 20, %64
    \qquad R_{\rm M}^\mathrm{max}(N_\perp = 32)\sim 
    81, 
    \qquad R_{\rm M}^\mathrm{max}(N_\perp = 64)\sim 326.
    \label{eq:conservative_res_limit_pencil}
\end{equation}
However, this bound is too optimistic since the discrete operators will deviate from the exact ones even before reaching the interparticle separation scale.
For example, for \textsc{Pencil}, the deviations from the exact operators appear when the ratio of the resolution scale to the feature scales
becomes $k_{\rm eff} / k_{\rm Ny} \simeq 0.7$ (see \textsc{Pencil} code documentation\footnote{\url{https://raw.githubusercontent.com/pencil-code/website/master/doc/manual.pdf}}),
or correspondingly $d_\mathrm{res}\simeq 1.2\lambda$.
Therefore, the maximal Reynolds number for \textsc{Pencil} can be $\simeq 2$ times smaller than the values above \Eq{eq:conservative_res_limit_pencil}.
In SPH, the actual resolution scale is larger than the interparticle distance, resulting in a smaller maximal magnetic Reynolds numbers.
For a quintic spline with a mean number of neighbours $N_{\rm neigh}\simeq 113$, then the kernel support radius is $3$ times the spacing between particles,
$d_\mathrm{res} = 3\lambda$, resulting in values $\simeq 9$ times smaller than in \Eq{eq:conservative_res_limit_pencil}, resulting in $R_{\rm M}^\mathrm{max}(64) = 36.2$.
For example, in \cite{STASYSZYN2015148}, the SPH runs for Roberts flow I were using $R_{\rm M}(50) = 200$, far more than the conservative limit.
Therefore, some resolution effects might change the observed growth and cause spurious effects. But, as was shown in Fig. 2 of \cite{SSKSB25}, significant growth rate deviations only started around $R_{\rm M}(64)\approx 200$.    
Although these bounds are very approximate, they illustrate the expected quadratic improvements with increasing spatial resolution
and give approximate conservative bounds on the region where numerical effects become significant; see \cite{SSKSB25} for a numerical demonstration for Roberts flow~I.

\section{Results}

In what follows, we investigate the dependency of the dynamo growth for the Roberts flows on the resistivity, or magnetic Reynolds number,
and assess the convergence of the codes using the scalar quantities of interest defined above and the magnetic field spatial distributions.

\subsection{Growth rate resistivity dependence for flows I, II, III, and IV}

\begin{figure}\begin{center}
\includegraphics[width=.55\columnwidth]{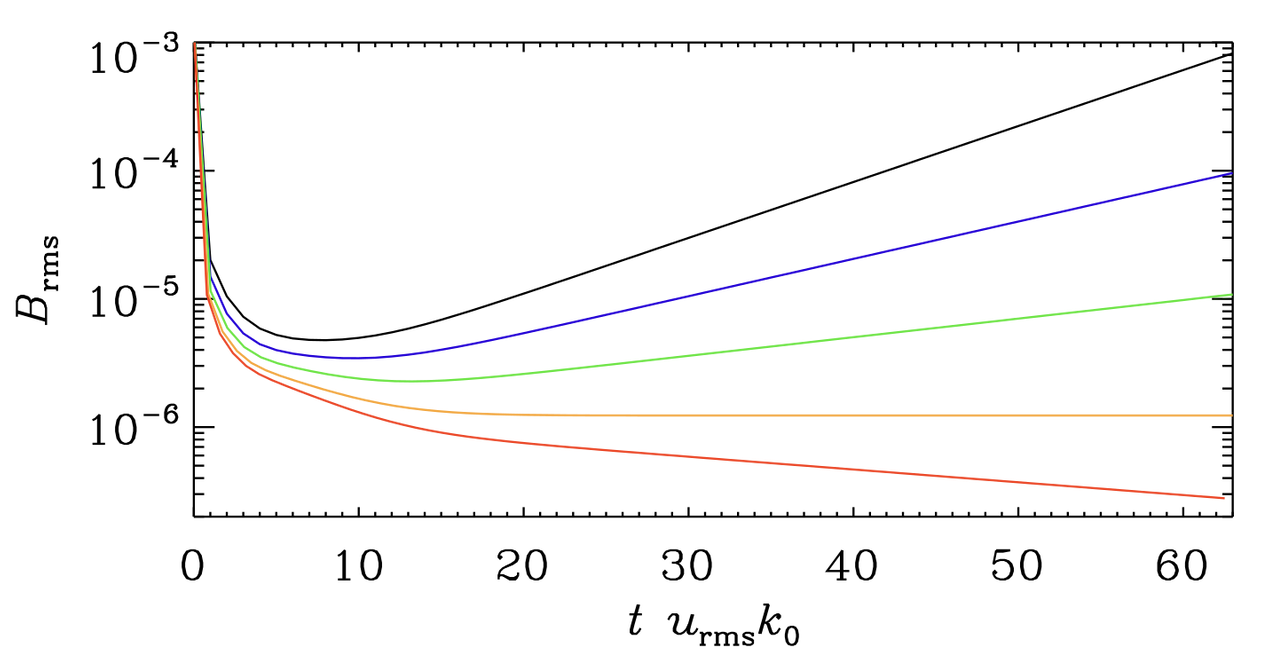}
\end{center}\caption[]{ Time dependence of $B_\mathrm{\rm rms}$ for different choices of resistivity ($\eta=0.1$ (black),
$\eta=0.127$ (blue), $\eta=0.154$ (green),
$\eta=0.1811$ (orange), and $\eta=0.2$ (red)) for \textsc{Pencil} 10th order scheme in a Roberts Flow I run near the dynamo onset point $\eta=0.1811$ with $w_0=\sqrt{2}v_0$ and $L_\perp=L_\|$.
}\label{pcomp}\end{figure}

Growing solutions for Roberts flows show an exponential increase or decrease of the magnetic field, as illustrated in \Fig{pcomp} for Roberts flow~I in a cubic domain.
The magnetic field decays initially because the random initial condition contains seeds for all modes, and only the fastest growing  (or the slowest decaying) contribute to $B_{\rm rms}$ after some time.
The value of the resistivity, $\eta$, above which an exponential growth is observed is the critical resistivity, $\eta_{\rm crit}$.
In what follows, we report the growth rate at late time as a function of the resistivity.
The time when growth rates commence depends on the flow and resistivity, because for larger resistivity growth rates tend to become smaller,
and as a consequence the mode has to grow longer to manifest itself from the initial conditions. 

\begin{figure}
    \centering
    \includegraphics[width=1\linewidth]{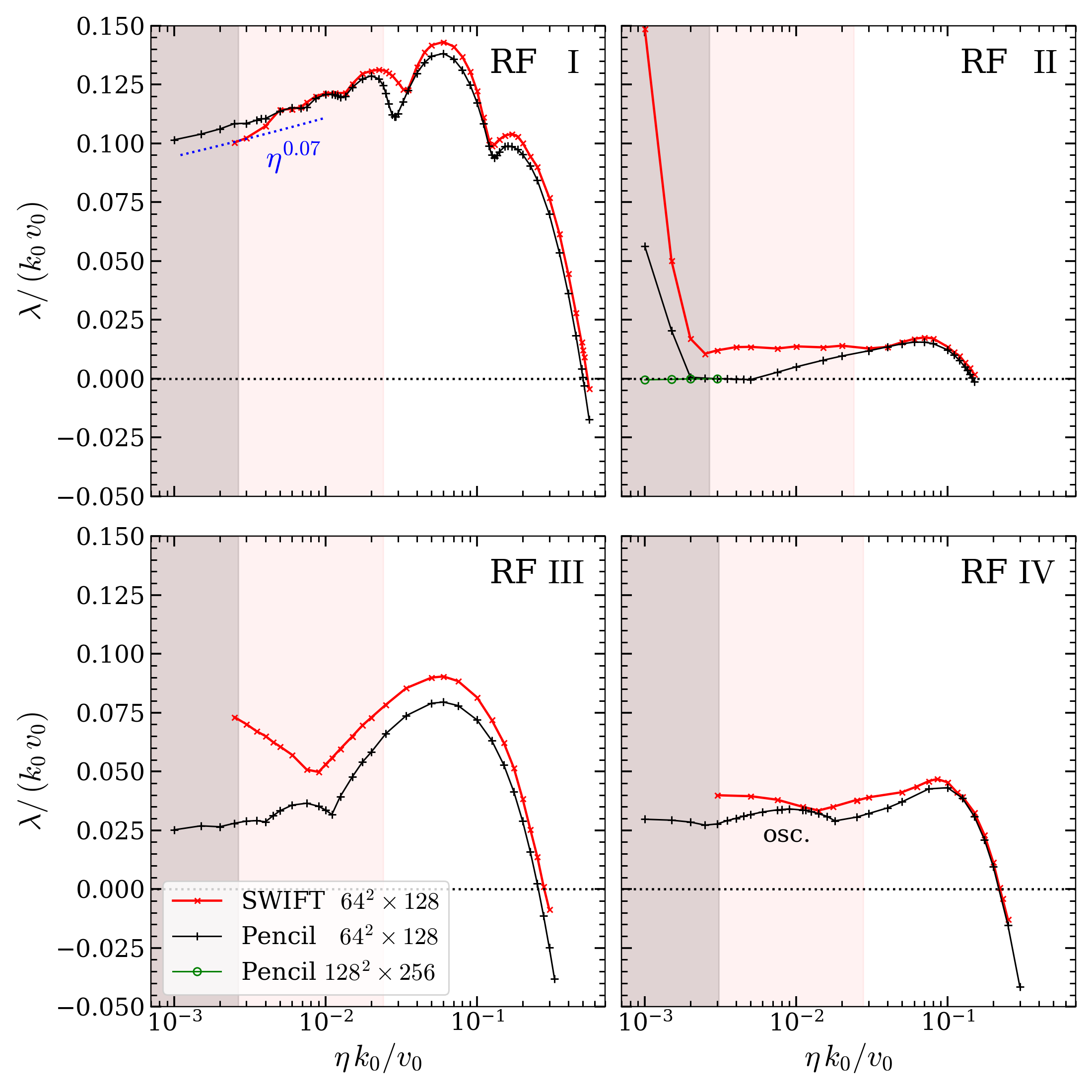}
    \caption{Growth rate dependence on resistivity for the 10th order \textsc{Pencil code} and \textsc{swift} codes. The runs were performed with the elongated box ($L_x=L_y=2\pi, L_z=4\pi$) with $64^2\times128$ resolution elements (particles or grid cells), and with the $ z$-direction velocity constant set to $w_0=v_0$. The black line with plus marker denotes \textsc{Pencil} code runs; the red line with cross marker denotes \textsc{swift} runs. We also provide $128^2\times256$ \textsc{Pencil} code runs outside the conservative limit, $ R_{\rm M} >R_{\rm M}^{\rm max}(\lambda)$ denoted with 2 green circular markers. Vertical shaded regions denote the conservative resolution limits, $R_{\rm M}^{\rm max}(\lambda)$ - for \textsc{Pencil} (gray band) (eq. \ref{eq:conservative_res_limit_pencil}) and $R_{\rm M}^{\rm max}(3\lambda)$ for \textsc{SWIFT} (red band). We also note the existence of a window with oscillating $B_{\rm rms}$ solutions for the flow IV} 
    \label{fig:roberts_flows_growth_rates}
\end{figure}

The four flows have different dynamo production mechanisms and solution properties. We set up a series of runs to study their growth rate dependence as a function of resistivity.
The resistivity is varied in the range $\eta \sim [10^{-3}, 0.6]$,
which roughly corresponds to the magnetic Reynolds number $R_{\rm M} = v_{\rm  rms}/(\eta k_0) \sim [1.4,1000]$ (depending on the flow).
To work with all flows in the same domain, we set up an elongated domain $L_\parallel  = 2 L_\perp $  with dimensions $L_\perp^2 \times L_\parallel$, $L_\perp=2\pi$. We choose $w_0=v_0$ to match the choice from \cite{Ntormousi_2023}.
In \Fig{fig:roberts_flows_growth_rates} we show a log-linear plot of the growth rate dependence on resistivity for all four flows for both codes. We used $64^2\times 128$ resolution elements (grid points or particles). The four flows are similar in that the non-zero growth starts once the magnetic Reynolds number becomes larger than the corresponding dynamo onset value. The onset is followed by a peak growth rate, which then decreases for smaller resistivity, demonstrating a slow dynamo behaviour.
First, we study the growth rate dependence with the \textsc{Pencil Code}.

For flow~I, we find the critical resistivity, $\eta_{\rm crit}\simeq 0.495$, which corresponds to $R_{\rm M}^{\rm crit} = v_{\rm rms} / (\eta k_0)\simeq 1.749$, where $v_{\rm rms}=0.866$ and $k_0 = 2\pi/L_\perp = 1.0$. In the limit of small resistivity, the growth rate behaves as $\lim_{\eta \rightarrow 0}\lambda \sim \eta^{0.07}$. 
In addition, the presence of sharp bends near $\eta\sim 1.3\cdot10^{-1}, \,2.8\cdot10^{-2}, \,1.3\cdot 10^{-2},7.4\cdot 10^{-3}$ indicates transitions between the fastest growing modes.

Flow II has only a finite window for dynamo action.
We find the window to be $\eta \in  [0.005,0.145]$ which corresponds to $R_{\rm M}^{\rm crit}\in [5.972,173.2]$.
A similar behaviour with finite windows has also been seen for ABC flow \citep[see e.g.][]{Galloway+Frisch86,BD13}.
We note that the growth rates at small resistivities $\eta \lesssim 0.05$ commence much more slowly for flow II compared to flow~I,
requiring a larger simulation time (up to $t=10^{3}$ for $R_{\rm M} \sim 10^2$). 

For flow III, we found a critical resistivity $\eta_{\rm crit}\simeq 0.255$ which corresponds to $R_{\rm M}^{\rm crit} \sim 3.43$. The flow reaches its peak growth rate around $\eta \simeq 0.05$ and grows more slowly for lower $\eta$. There are mode transitions around $\eta \simeq 1.1\cdot 10^{-2}, \, 4.0\cdot 10^{-3}$.
For lower values of the resistivity, the flow requires a long time $t\simeq 500$ for the growing modes to fully develop.

For flow IV, the critical resistivity is $\eta_{\rm crit}\simeq 0.220$ and $R_{\rm M}^{\rm crit} = v_{\rm rms}/(\eta k_0) \sim 4.54$, where $v_{\rm rms} = 1.0$.
The growth rates reach a maximum around $\eta \simeq 0.1$ followed by a decrease for smaller resistivity.
Surprisingly, there is a window $\eta \in [ 2.8\cdot 10^{-3},1.9\cdot 10^{-2}]$ or $R_{\rm M} \in [53,357]$ where, unlike other flows, the growing solutions are oscillatory, similar to the growth observed in the ABC flow. The values of growth rates and frequencies are given in \Tab{TflowIV}.

Both \textsc{Pencil} and \textsc{SWIFT} recover all Roberts flow features for magnetic Reynolds numbers smaller than their conservative limit defined in \Sec{sec:spacial_resolution_limits}
and have relatively small deviations from each other. \textsc{SWIFT} produces larger growth rates than \textsc{Pencil} across all resistivities by roughly the same absolute amount depending on the flow.
This difference also lowers the critical magnetic Reynolds numbers for \textsc{SWIFT} compared to the \textsc{Pencil} code values as shown in \Tab{Tcrit}.
In addition, some mode transitions, such as the second transition for flow I, which in \textsc{Pencil} happens at $\eta \simeq 2.8\cdot 10^{-2}$, are shifted to $\eta \simeq 3.5\cdot 10^{-2}$. However, we note that this transition is close to the conservative resolution limit.
For magnetic Reynolds numbers larger than the respective conservative limit, the performance depends on the flow.
For flows I, III, and IV, significant deviations between the codes appear after $R_{\rm M}(3\lambda)$, closer to $R_{\rm M}(\lambda)$. The dynamo window in Roberts flow II is not fully covered by \textsc{SWIFT} at $64^2\times 128$ resolution; significant growth rate overestimation appears after crossing $R_{\rm M}(3\lambda)$.
For flow~II, \textsc{SWIFT} produces strongly growing solutions when $R_\mathrm{M}>R_\mathrm{M}(\lambda)$.
Interestingly, the \textsc{Pencil} code also produces very strongly growing solutions for $R_{\rm M}>R_{\rm M}^\mathrm{max}(\lambda)$. But higher-resolution $128^2\times 256$ \textsc{Pencil} runs suggest that the observed growth is unphysical.
Further, in \Sec{sec:convergence_scalars}, we show for Roberts flow~I that the quantities defined in \Sec{sec:observable_characteristic_quantities} converge with resolution for the two codes.

%%%%%%%%%%%%%%%%%%%%%%%%%%%%%%%%

%FIG1

\begin{table}\caption{
Summary of the critical resistivities and magnetic Reynolds numbers obtained with \textsc{Pencil} and \textsc{SWIFT} codes for $64^2\times 128$ resolution runs of the the Roberts flows~I--IV
}\vspace{12pt}\centerline{\begin{tabular}{lccccccc}
  & Scheme & I &  II &  III &  IV \\
\hline
$\eta_{\rm crit}$ & \textsc{Pencil 10th}  & 0.495 &   0.145  & 0.255 &  0.220 \\
 & \textsc{SWIFT}  & 0.537 &  0.155  & 0.279 &  0.224 \\

 $R_M^{\rm crit}$ & \textsc{Pencil 10th}  & 1.749 &  5.972  & 3.396 &  4.545 \\
 & \textsc{SWIFT}  & 1.613 &  5.587  & 3.104 &  4.464 \\

\label{Tcrit}\end{tabular}}\end{table}

\subsection{Impact of box geometry and $w_0$ for Roberts flow~I}
The growth rate dependence from \Fig{fig:roberts_flows_growth_rates}, described above, is also affected by the shape of the periodic domain and the magnitude of the velocity in the $z$ direction.
Below, we investigate how this dependence changes under the domain elongation from $L_\|=L_\perp$ to $L_\|=2L_\perp$ and the change in the $z$ velocity from $w_0 = \sqrt{2}v_0$ to $w_0 = v_0$.

\begin{figure}\begin{center}
\includegraphics[width=0.7\columnwidth]{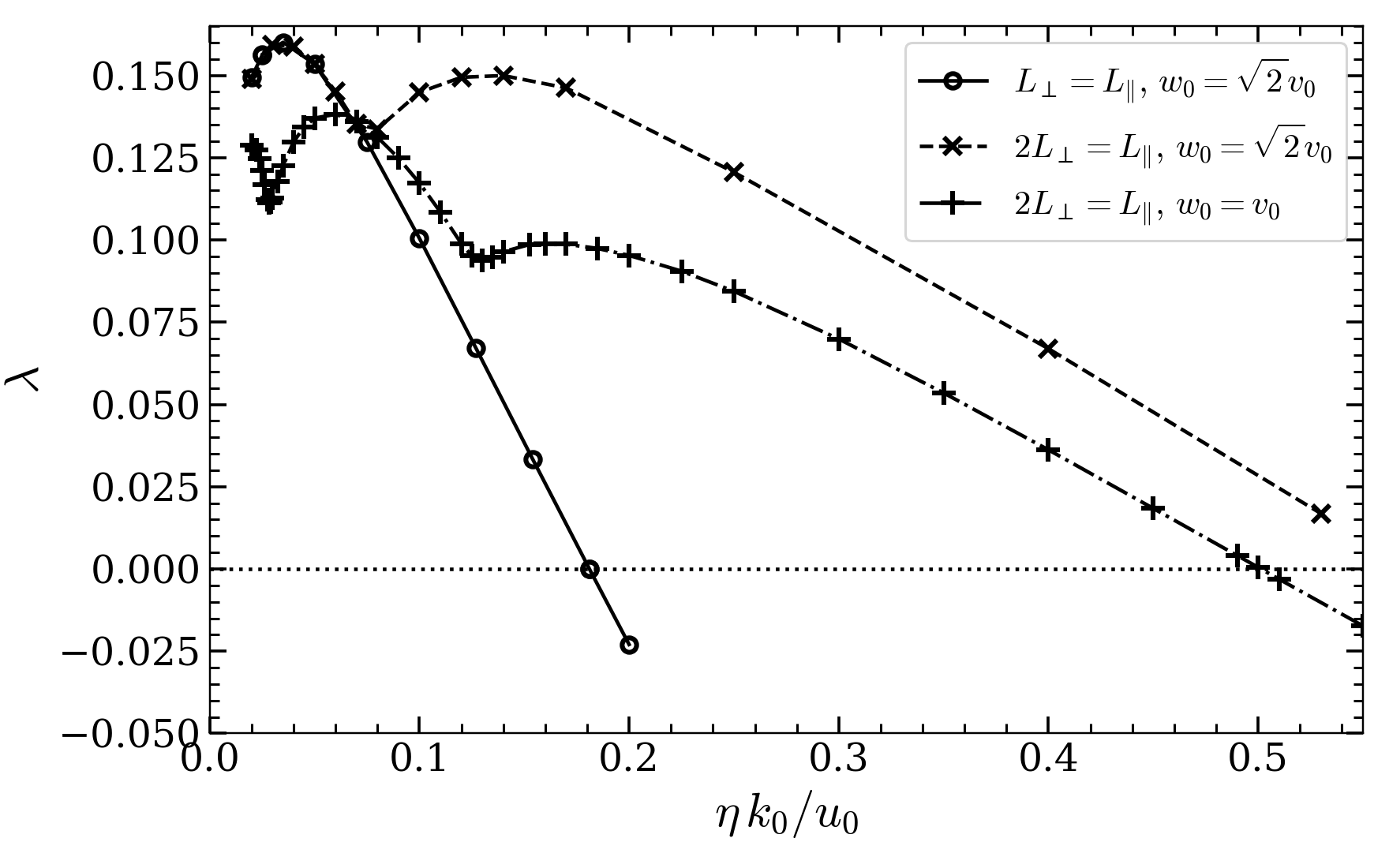}
\end{center}\caption[]{
Illustration of the change in the Roberts flow I growth rate in response to changes in the domain size and the velocity in the $z$ direction, $w_0$. Note that, unlike \Fig{fig:roberts_flows_growth_rates}, the x-axis is not logarithmic. For the cubic domain (solid line), the dynamo onset happens around $\eta = 0.1811$. For the elongated domain $L_\| = 2L_\perp$ with the same $w_0$, an additional mode with $k=0.5$ can be excited, and as a result the critical resistivity increases to around $\eta \simeq 0.57$. Changing the velocity in the $z$ direction, $w_0$, for the elongated domain results in a change in the growth rate dependence on resistivity for each mode, and in shifted mode transitions, as well as a shifted critical resistivity to $\eta_{\rm crit} = 0.495$, but it does not introduce new modes. The critical Reynolds number, though, stays roughly the same at $R_{\rm M} \simeq 1.747$ under the change of $w_0$.}\label{fig:Domain_and_w0_vs_growth_rate}\end{figure}

In many applications for Roberts flow~I, $w_0\equiv v_0 \sqrt{2}$ was chosen
\citep{BS05, BC20}, while in the numerical solutions presented by \cite{Rhei+14} and  \cite{Ntormousi_2023},
$w_0=v_0$ was used.
In \Fig{fig:Domain_and_w0_vs_growth_rate}, we explore, using the \textsc{Pencil} code only, how this choice impacts the growth rate dependence on the resistivity.
The computational domain has the size $L_\perp\times L_\perp\times L_\|$.
For the case of cubic domain $L_\perp= L_\|$ and the choice $w_0\equiv v_0 \sqrt{2}$, the dynamo onset occurs at $\eta\approx0.1811\,v_0/k_0$.
This means that the critical magnetic Reynolds number is $R_{\rm M}=v_\mathrm{rms}/\eta k_0\approx5.52$. 

For the elongated domain with the same choice $w_0\equiv v_0 \sqrt{2}$, 
a mode with $k_z=0.5$ can appear, which leads to an increase in the critical resistivity to  $\eta\approx0.5725\,v_0/k_0$ or a decrease in the magnetic Reynolds number
to $R_{\rm M}\approx1.747$. For flows II and III, the impact is more dramatic because there is no dynamo in the cubic domain.

Changing $w_0$ from $\sqrt{2}v_0$ to $v_0$ (and thus a change in $v_{\rm rms}$ from $1.0$ to $0.866$) for the elongated domain, $2L_\perp = L_\|$, changes the behaviour for each mode with resistivity, and decreases the critical resistivity to $\eta_{\rm crit} = 0.495$. However, the critical magnetic Reynolds number stays roughly the same, $R_{\rm M} = v_{\rm rms}/(\eta k_0)\approx1.749$. There are also some changes in the growth rate magnitude and the positions of the mode transition points. The first transition for the elongated box and $w_0=v_0$ the transition happens at $\eta \simeq 1.3\cdot 10^{-1}$ ($R_{\rm M} \simeq 6.67$) while for another choice $w_0=\sqrt{2}v_0$ the transition shifts to $\eta \simeq 7.5 \cdot 10^{-2}$ ($R_{\rm M} \simeq 13.3$)

%FIG2

\begin{table}\caption{
Growth rates and frequencies for Roberts flow IV oscillatory growth window denoted as 'osc.' in \Fig{fig:roberts_flows_growth_rates}.
}\vspace{12pt}\centerline{\begin{tabular}{lccccccc} 
$\eta k_0/v_0$       & $0.0030$ & $0.0050$ & $0.0075$ & $0.0110$ & $0.0125$ & $0.0140$ & $0.0180$ \\
\hline
$\lambda / (v_0k_0)$ & $0.0277$ & $0.0317$ & $0.0337$ & $0.0336$ & $0.0330$ & $0.0321$ & $0.0290$ \\
$\omega_{\rm mod} /  (v_0k_0)$ & $0.0797$ & $0.0975$ & $0.1172$ & $0.1397$ & $0.1480$ & $0.1561$ & $0.1730$ \\
\label{TflowIV}
\end{tabular}}\end{table}

%FIG5
\begin{figure}
\begin{center}
\includegraphics[width=0.9\columnwidth]{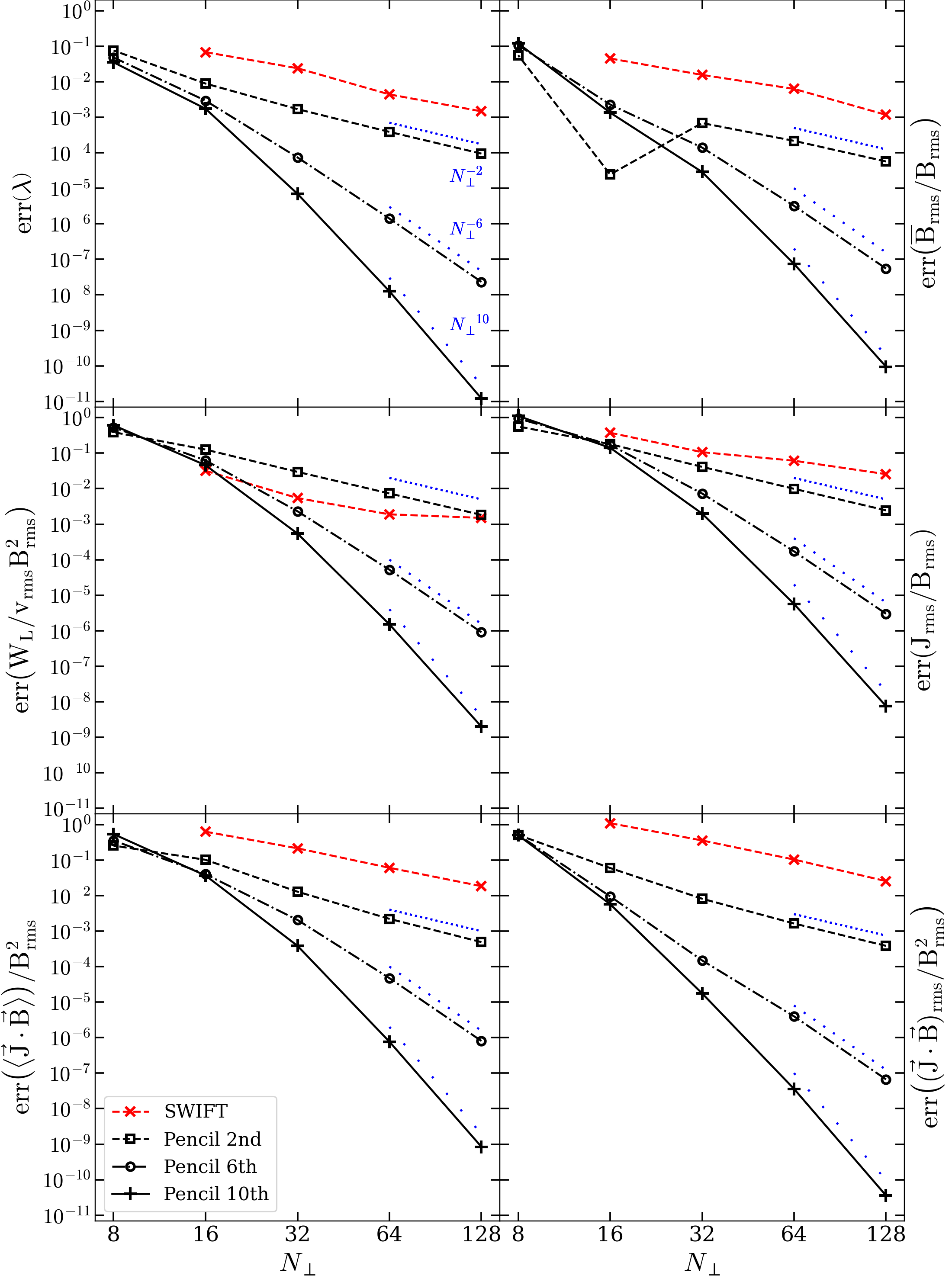}
\end{center}
\caption[]{
Convergence properties of different quantities for second-, sixth-, and
tenth-order \textsc{Pencil} schemes shown as dashed, dash-dot, and solid lines with square, circular and plus markers, respectively. Convergence for \textsc{SWIFT} is shown as a red dashed line with a cross marker. Convergence is shown for \textsc{Pencil} with respect to its highest-resolution $256^2\times 512$ 10th-order result, and for \textsc{SWIFT} with respect to its $256^2\times 512$ run.
In each panel, the short dotted, long dotted, and longer dotted lines denote
$\propto1/N_\perp^2$, $1/N_\perp^6$, and $1/N_\perp^{10}$ slopes.
\textsc{Pencil} schemes demonstrate regions of convergence that corresponds to the order of the scheme, that is 2nd order converges as $N_\perp^{-2}$, 6th order as $N_\perp^{-6}$, 10th order as $N_\perp^{-10}$, while \textsc{SWIFT} converges as 2nd order \textsc{Pencil} code but with effectively 3 times worse linear resolution. We note that \textsc{SWIFT} shows deviations from $N_\perp^{-2}$ convergence only for the work done against the Lorentz force.  
}\label{p5em2}\end{figure}

\subsection{Impact of resistivity and flow choice on the quantities of interest}
To get an idea of how the characteristic quantities defined in section \ref{sec:observable_characteristic_quantities} depend on both $\eta$ and the flow choice, we performed 4 low-resolution runs with different resistivities for flows I and II
in \Tabs{Tresults}{Tresults2}. 
The tables contain four columns for runs for each resistivity value. The rows contain the values of plane averaged field $\tilde B_{\rm rms} / B_{\rm rms}$,
the magnetic Taylor microscale wavenumber, $J_{\rm rms}/B_{\rm rms}$, the mean magnetic helicity density $\langle \bm{A}\cdot \bm{B}\rangle/B_{\rm rms}^2$,
the work done against the Lorentz force $W_\mathrm{L}$ normalised with $v_{\rm rms} J_{\rm rms}B_{\rm rms}$ and various averages of $\bm{J}$ and $\bm{B}$ products:
$\langle \bm{J} \cdot \bm{B} \rangle $, $\langle (\bm{J} \cdot \bm{B})^2 \rangle^{1/2}$,
$\langle (\bm{J} \times \bm{B})^2 \rangle^{1/2}$, and $\langle \bm{J}^2 \bm{B}^2 \rangle$ normalised with appropriate powers of $J_{\rm rms}B_{\rm rms}$. 
As seen before, for flow~I, $\lambda$ decreases only very slowly for resistivities smaller than 
$\eta k_0/v_0<0.05$.
We also see that the fractional mean field, $\meanBrms/\Brms$, decreases
slowly (approximately $\propto\eta^{1/4}$) with decreasing $\eta$, while
the magnetic Taylor microscale wavenumber increases like $\eta^{-1/2}$
and the normalised work done against the Lorentz force decreases like
$\eta^{1/2}$.

With regards to ideal MHD limit of $\eta\rightarrow0$, the dependencies of the characteristic quantities of dynamos on $\eta$,
as seen above, are typical of both slow and (supposedly) fast dynamos.
In the absence of diffusion (the strict case $\eta=0$, as opposed to the limit
$\eta\to0$), no eigenfunction is possible and the scale of the magnetic
field becomes smaller as time goes on.
The resulting rms magnetic field can still grow exponentially in time,
but the spatial structure of the magnetic field continues to become
finer with time.
As a consequence, the magnetic Taylor microscale and some other similar
ratios that were constant for a proper eigenfunction remain time-dependent
and can therefore not be eigenfunctions.
We recall, however, that such a solution is not a fast dynamo in the sense that
no eigenfunction exists.

In this connection, we recall that in the resistively accomplished
marginal steady state, $\bra{\JJ\cdot\BB}=0$; see \cite{Bra01} and \cite{BS20}.
This is indeed confirmed for Roberts flow~I, where $\bra{\JJ\cdot\BB}=0$
is found when $\eta\approx0.1811\,v_0/k_0$ is equal to the marginal value;
see \Tab{Tresults}.
For Roberts flow~II, on the other hand, we find $\bra{\JJ\cdot\BB}=0$
for all values of $\eta$; see \Tab{Tresults2}.
This is because the flow has pointwise vanishing helicity, so there cannot
be any source of magnetic or current helicity.

We have not listed here the values of the current helicity of the mean
magnetic field, i.e., of $\meanJJ\cdot\meanBB$.
This is because in the helical case, i.e., for Roberts flow~I, we always
have $\meanJJ\cdot\meanBB/\meanJrms\meanBrms=-1$ owing to the fact that
the kinetic helicity of the flow is positive \citep{Bra01}.
This is also confirmed by the present results, but such a check should
also be part of any other code benchmark.

\subsection{Roberts flow high resolution reference runs}
An important aspect of the numerical analysis of Roberts flows is the convergence of all results with numerical resolution to the true values a continuous system would have.
Because we do not have analytical solutions, we choose to fix the reference value to be equal to the highest resolution runs performed for the system.
Below we discuss these reference runs, show the reference values.
In \Tabss{Tresults}{TConvFit} and choose suitable parameters for the convergence study for both \textsc{Pencil} and \textsc{SWIFT} codes. 
We choose flow~I with $\eta = 0.05$ as a reference for the study of the convergence properties for several reasons:
it roughly corresponds to the peak growth rate and the onset of slow dynamo behaviour, and the larger the growth rate is, the smaller it takes for the growth to commence.
Furthermore, the point is far from mode transitions and therefore, contribution to the growth rate measurements from other modes will be smaller;
studying convergence for smaller resistivity is computationally more expensive and shifts the bound on the lowest resolution run
to higher resolutions because for smaller resistivity the typical scales of feature is smaller and requires more resolution.
Also, flow~I produces significantly larger growth rates than the other flows, as can be seen for example for the case $w_0 = v_0$ in \Fig{fig:roberts_flows_growth_rates}.
In addition, instead of $w_0 = v_0$, we choose $w_0 = \sqrt{2}v_0$ also in part because the resulting growth rates are larger, as can be seen from \Fig{fig:Domain_and_w0_vs_growth_rate},
and because this parameter choice was used in previous studies \citep[e.g.][]{Tilgner_2008,SSKSB25} and will produce the same growth rate for both elongated and cubic geometries. 

\begin{figure}
    \centering
    \includegraphics[width=0.65\linewidth]{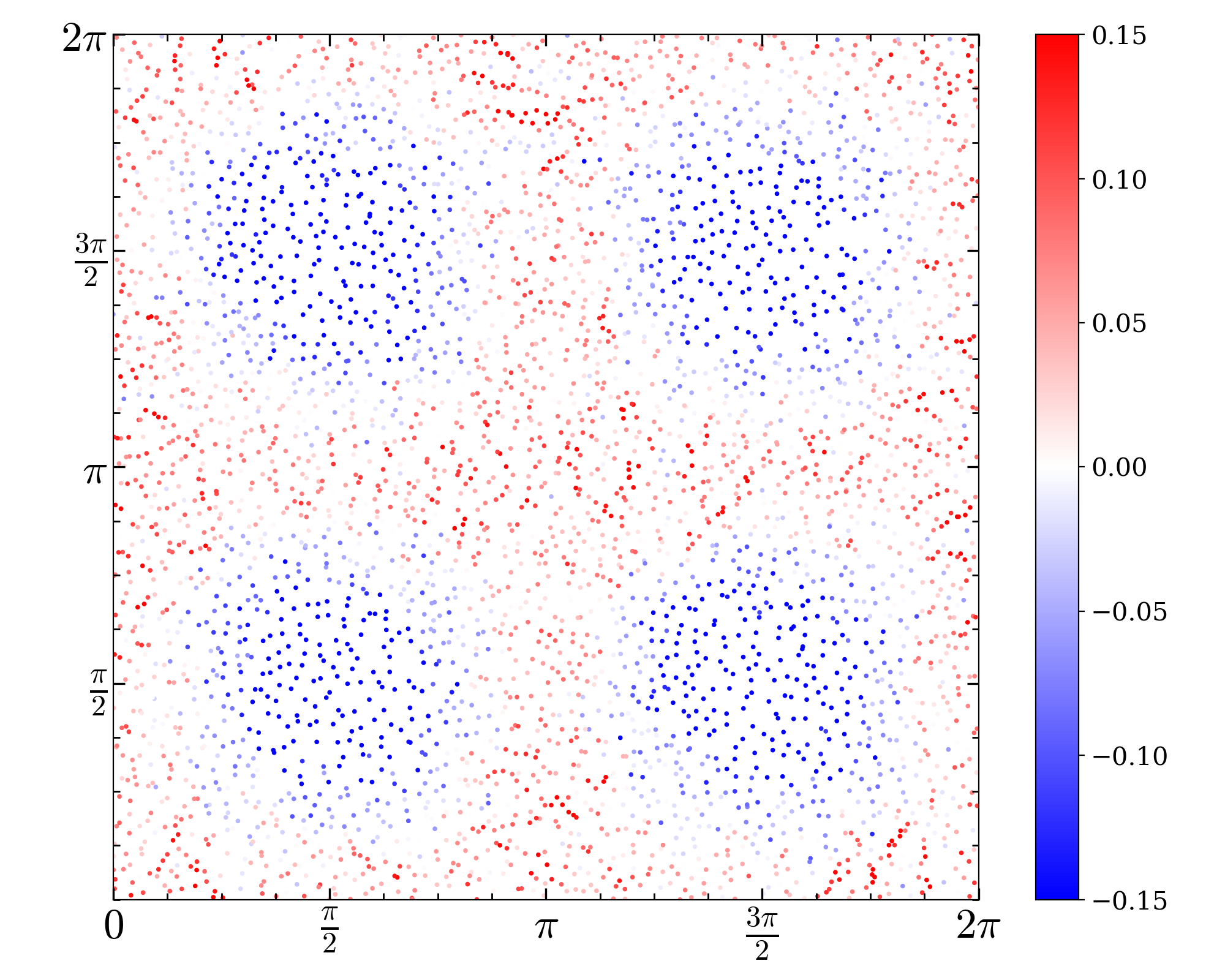}
    \caption{Slab of the \textsc{SWIFT} $64^2\times128$ Roberts flow 1 run at $z=2\pi$ of height of one interparticle spacing $d = (m/\rho)^{1/3}$ at $t=100$.
Colour shows the density fluctuation magnitude. Over the simulation volume, it mostly stays below $15 \%$. Although initially particles were arranged in a relaxed glass-like configuration, shear will distort the particle arrangement. The effect depends on the sound speed magnitude.}
    \label{fig:SWIFT_particle_slab}
\end{figure}

We ran this setup with \textsc{Pencil} at  $256^2\times 512$ resolution (for the discussion about the choice of  $N_\|$ see Appendix~\ref{appx:Npar}).
As initial conditions, we choose to excite the growing mode directly using \Eq{equation:initial_condition} instead of employing a random vector potential.
The latter choice would excite several modes simultaneously, and in order to measure the growth rate more accurately,
with relative errors less than $\delta \lambda/\lambda \lesssim 10^{-11}$, the required time would be $t_\mathrm{max}\sim 10^{3}$,
while sinusoidal initial conditions suppress all other growing modes to machine precision, and require only $t_\mathrm{max}\sim10^2$.
We choose the same $256^2 \times 512$ spatial resolution for \textsc{SWIFT}, that is, with the same particle spacing in the $z$ direction as in the $xy$ plane.
The comparison of the run results between \textsc{Pencil} and \textsc{SWIFT} is shown in \Tab{TresultsPencilSwift}.
The table contains two columns for \textsc{Pencil}, one with quantities of interest computed with internal gradient operators as in \Tabs{Tresults}{Tresults2},
while the other employs the same reconstruction, described in \Sec{sec:code_comparison}, used for both \textsc{SWIFT} and \textsc{Pencil} to ensure there are no significant systematic effects.
The obtained growth rates are $\sim 10\%$ larger than for the same \textsc{Pencil} 10th order run.
One possible source of such errors for SPH operators is the inhomogeneous and anisotropic neighbour distribution.
Figure~\ref{fig:SWIFT_particle_slab} shows the particle distribution in a slab of a thickness equal to the interparticle spacing in that plane.
Individual particles are shown as dots, with colour denoting density fluctuations. We observe density fluctuations with magnitude $\sim 10\%$ which follow the structure of the flows: inside vortices the density is lower than in shear zones.
In addition, the local particle distribution inside vortices is less irregular than within the shear zones, where the magnetic field is stretched and amplified.
In Roberts flows, the growth rate follows from an interplay between the stretching and resistive terms.
Such density fluctuations and particle distribution irregularities can decrease the accuracy of SPH derivative operators, particularly operators for the velocity gradient, relevant for the stretching term, and the SPH Laplacian operator, which enters the resistivity diffusion term  \citep{Price_2012}.
Nevertheless, in \Tab{TresultsPencilSwift} we see good agreement within $\sim 10\%$, between the values obtained in \textsc{Pencil} and \textsc{SWIFT}.

\begin{table}\caption{
Summary of quantities (\Sec{sec:observable_characteristic_quantities}) for flow~I using four values of $\eta k_0/v_0$ with $w_0=\sqrt{2}\, v_0$ and $32^3$ resolution runs.
We note that for such a parameter choice, the cubic box and the elongated box result in the same growth rate, as shown in \Fig{fig:Domain_and_w0_vs_growth_rate}.
}\vspace{12pt}\centerline{\begin{tabular}{lccccccc}

$\eta k_0/v_0                           $ &   0.1811 &   0.0500 &    0.0010 &    0.0002 \\
\hline
$\lambda/v_0 k_0                        $ &   0.0000 &   0.1534 &    0.114 &    0.0745 \\
$\meanBrms/\Brms                        $ &   0.6685 &   0.5868 &    0.257 &    0.2023 \\
$\mu_0\Jrms/k_1\Brms                    $ &   1.6691 &   2.3607 &    15.08 &    36.089 \\
$\bra{\JJ\cdot\BB}/\Jrms\Brms           $ &   0.0000 &   0.2594 &    0.057 &    0.034 \\
$\bra{(\JJ\cdot\BB)^2}^{1/2}/\Jrms\Brms $ &   0.6545 &   0.9064 &    0.439 &    0.304 \\
$\bra{(\JJ\times\BB)^2}^{1/2}/\Jrms\Brms$ &   0.8391 &   1.1769 &    3.836 &    4.298 \\
$-W_\mathrm{L}/\urms\Jrms\Brms          $ &   0.3023 &   0.1830 &    0.023 &    0.010 \\
$\bra{\JJ^2\BB^2}/\Jrms^2\Brms^2        $ &   1.1325 &   2.2065 &   14.908 &   18.562 \\
$-\bra{\AAA\cdot\BB}/\Brms^2            $ &   0.3496 &   0.1996 &    0.008 &    0.0034 \\
\label{Tresults}\end{tabular}}\end{table}

\begin{table}\caption{
Results for Roberts flow II employing the same setup as in \Tab{Tresults}.
}\vspace{12pt}\centerline{\begin{tabular}{lccccccccc}

$\eta k_0/v_0                           $ &   0.1399 &   0.1000 &    0.0500 &    0.002 \\
\hline

$\lambda/v_0 k_0                        $ &   0.0000 &   0.0105 &    0.010 &    0.001 \\
$\meanBrms/\Brms                        $ &   0.4030 &   0.3567 &    0.282 &    0.000 \\
$\mu_0\Jrms/k_1\Brms                    $ &   1.8201 &   2.0245 &     2.55 &    10.23 \\
$\bra{\JJ\cdot\BB}/\Jrms\Brms           $ &  -0.0009 &   0.0002 &   -0.000 &   -0.000 \\
$\bra{(\JJ\cdot\BB)^2}^{1/2}/\Jrms\Brms $ &   0.5871 &   0.6277 &    0.696 &    1.035 \\
$\bra{(\JJ\times\BB)^2}^{1/2}/\Jrms\Brms$ &   0.9024 &   0.8775 &    0.848 &    1.639 \\
$-W_\mathrm{L}/\urms\Jrms\Brms          $ &   0.2546 &   0.2076 &    0.131 &    0.021 \\
$\bra{\JJ^2\BB^2}/\Jrms^2\Brms^2        $ &   1.1676 &   1.1732 &    1.214 &    3.777 \\
$-\bra{\AAA\cdot\BB}/\Brms^2            $ &  -0.0001 &   0.0007 &   -0.000 &   -0.000 \\
\label{Tresults2}\end{tabular}}\end{table}

\begin{table}\caption{
Summary of the quantities (sec. \ref{sec:observable_characteristic_quantities}) for flow~I with $\eta k_0/v_0 = 0.05$ at $N_\perp=256$ line element resolution with $w_0=\sqrt{2}\, v_0$. We set $k_1=1.0$ - for the wavenumber of the fastest mode at the chosen resistivity. We provide the results for the \textsc{Pencil} 10th-order scheme run at $256^2\times 512$ (up to 12 digits, first and second column; see also Appendix \ref{appx:convergence_fit}), corresponding to \textsc{SWIFT} results. The first column was obtained directly from the grid data, while the second and the third columns were obtained with decomposition in Fourier space (see sec.  \ref{sec:code_comparison}). The difference between the first and the second column starts at the 8th digit, which is much less than the levels of resolution error in \textsc{SWIFT}, indicating that the decomposition used for \textsc{SWIFT} is not responsible for the observed difference in the quantities of interest. }\vspace{12pt}\centerline{\begin{tabular}{lccc}
quantity                                  &   \textsc{Pencil} 10th  &  \textsc{Pencil} 10th FT & \textsc{SWIFT}  \\
\hline
$\lambda/v_0 k_0                        $ &     0.153398401832 &  - &  0.1642 \\
$\meanBrms/\Brms                        $ &     0.586753221004 &  0.58675322 &  0.5906 \\
$\mu_0\Jrms/k_1\Brms                    $ &     2.360716135475 &  2.36071600 &  2.3385 \\
$\bra{\JJ\cdot\BB}/\Jrms\Brms           $ &     0.259392758997 &  0.25939274 &  0.2575 \\
$\bra{(\JJ\cdot\BB)^2}^{1/2}/\Jrms\Brms $ &     0.906396536089 &  0.90639641 &  0.8960 \\
$\bra{(\JJ\times\BB)^2}^{1/2}/\Jrms\Brms$ &     1.176856767525 &  1.17685661 &  1.1607 \\
$-W_\mathrm{L}/\urms\Jrms\Brms          $ &     0.183015411702 &  0.18301542 &  0.1842 \\
$\bra{\JJ^2\BB^2}/\Jrms^2\Brms^2        $ &     2.206546531903 &  2.20654595 &  2.1501 \\
$-\bra{\AAA\cdot\BB}/\Brms^2$             &     0.199595518688 &  0.19959552 &  0.2029  \\
\label{TresultsPencilSwift}\end{tabular}}\end{table}

% Original values 16 digits
% \vspace{12pt}\centerline{\begin{tabular}{lccc}
% quantity                                  &   Pencil 10th  &  Pencil 10th FT &  \textsc{SWIFT}  \\
% $\lambda/v_0 k_0                        $ &     0.1533984018321584 &  - &  0.1642 \\
% $\meanBrms/\Brms                        $ &     0.5867532210040555 &  0.5867532240552119 &  0.5906 \\
% $\mu_0\Jrms/k_1\Brms                    $ &     2.3607161354748056 &  2.3607160090020485 &  2.3385 \\
% $\bra{\JJ\cdot\BB}/\Jrms\Brms           $ &     0.2593927589969672 &  0.25939274856052646 &  0.2575 \\
% $\bra{(\JJ\cdot\BB)^2}^{1/2}/\Jrms\Brms $ &     0.9063965360888537 &  0.9063964198025147 &  0.8960 \\
% $\bra{(\JJ\times\BB)^2}^{1/2}/\Jrms\Brms$ &     1.1768567675247059 &  1.1768566139845957 &  1.1607 \\
% %$W_\mathrm{L}/\urms\Jrms\Brms           $ &     0.1830 & -0.1757 & -0.1842 \\
% $-W_\mathrm{L}/\urms\Jrms\Brms          $ &     0.1830154117023487 &  0.1830154210436984 &  0.1842 \\
% $\bra{\JJ^2\BB^2}/\Jrms^2\Brms^2        $ &     2.2065465319025730 &  2.2065459597101045 &  2.1501 \\
% %$\bra{\AAA\cdot\BB}/\Brms^2$              &     0.4320 & -0.1994 & -0.2029  \\
% $-\bra{\AAA\cdot\BB}/\Brms^2$             &     0.1995955186880222 &  0.19959552194436647 &  0.2029  \\
% \label{TresultsPencilSwift}\end{tabular}}\end{table}

%FIG5

\subsection{Convergence behaviour} \label{sec:convergence_scalars}

Another important aspect is how fast the results converge with resolution increase to the reference values.
After choosing the parameters for the reference run above for both \textsc{Pencil} and \textsc{SWIFT},
we look at the convergence power of the deviations for 6 quantities of interest with respect to reference values.
We investigate the convergence for the 2nd-, 6th-, and 10th-order \textsc{Pencil} schemes and for the \textsc{SWIFT} code.
This is shown in \Fig{p5em2}.

% In the following, we present quantitative results for characterizing
% the eigensolutions.

Generally, the errors for the quantities of interest, defined in \Eq{eq:error_def}, decrease with increasing $N_\perp$ as $1/N_\perp^2$
for the second-order scheme, $1/N_\perp^6$ for the sixth-order scheme,
and $1/N_\perp^{10}$ for the tenth-order scheme.
Starting at $64^2\times 128$ resolution, all quantities of interest follow the slopes of the corresponding schemes.
The \textsc{swift} code converges to its highest-resolution run ($256^2\times 512$) as $1/N_\perp^2$,
in a similar way to the second-order \textsc{Pencil} scheme but with effectively $\simeq 3$ times smaller linear spatial resolution.
This result can be interpreted as discussed above in \Sec{sec:spacial_resolution_limits},
because the resolution scale in SPH is larger than the interparticle spacing and can be conservatively estimated as a kernel cutoff scale, which is 3 times larger than the interparticle spacing.
The $1/N_\perp^2$ convergence agrees with the previously studied Roberts flow~I convergence in a cubic domain \citep{SSKSB25}
for the critical resistivity; it was found that the value converges as $\sim h^2$, which is the same as was found in this study because $h\sim N_\perp^{-1}$.
Although \textsc{SWIFT} has formally first-order accurate derivative operators,
such a $N_\perp^{-2} \sim h^2$ convergence confirms the mechanism discussed by \cite{Price_2012},
where the first-order errors in the pressure gradient term rearrange particles until the errors are minimised,
resulting in improvements in accuracy.
Given such a reliance on particle distribution,
one might expect $\sim N_\perp^{-1}$ scaling instead in more general astrophysical MHD applications.
In \Tab{TresultsPencilSwift}, we note that \textsc{swift} runs converge to larger growth rates than the \textsc{Pencil} 10th-order scheme for the same number of resolution elements.
The deviation of \textsc{SWIFT} growth rates compared to \textsc{Pencil} is $\delta\lambda/\lambda\sim 10^{-1}$,
which is larger than the relative difference between $128^2\times256$ and the $256^2\times512$ reference runs with \textsc{SWIFT}, $\delta\lambda/\lambda\sim 10^{-3}$,
indicates that there are systematic resolution-independent effects in \textsc{SWIFT} runs, responsible for the growth rate overestimation.  One example could be the influence of the time step choice on the growth rates. The choice of the time step is discussed in Appendix  \ref{appx:SWIFTdt}.

We also see some departures from the convergent behaviour for \textsc{SWIFT}. Particularly notable is the fact that the work against the Lorentz force, $W_\mathrm{L}/(v_{\rm rms} \Brms^2)$, no longer converges
with increasing $N_\perp$, while others continue to converge.
The linear fits to the convergence plots, the results are shown in Appendix \ref{appx:convergence_fit}.

\subsection {Time-independent magnetic field patterns for Roberts flows I, II, III, and IV}

In addition to the values for the scalar quantities of interest, the spatial magnetic field distribution of the growing eigensolutions can provide a unique,
feature-rich fingerprint characteristic for each flow.
As discussed in \Sec{sec:growing_solutions}, because all four flows are independent of the $z$ direction, the growing solutions
have a simple plane wave dependence on $z$, $\propto e^{ik_z z}$, and all the other information can be compacted in a 2D map of the complex magnetic field,
or equivalently, 6 real maps described in \Eq{ModulusPhase}:
three magnetic field amplitudes, $r_1,r_2,r_3$, for each spatial direction;
and three phases $\theta_1,\theta_2,\theta_3$ that show how does the wave $e^{ik_z z}$ experience a phase shift in the $z$ direction for each point in the $xy$ plane.
We show these 6 density maps for each flow.
However, as discussed in \Sec{sec:time_independent_patterns}, there is an additional complication for flow II because the growing solution is composed of two modes
of perpendicular polarisation (one has a non-zero plane-averaged $\tilde B_x$, while the other has a non-zero plane-averaged $\tilde B_y$) and under such a visualisation,
they will interfere and produce time-dependent images.
To avoid this problem, we use simple sine-wave initial conditions from \Eq{equation:initial_condition} to excite only one mode.
With this, we report the resulting unique time-independent magnetic field patterns characterising the growing 3D eigensolutions for each Roberts flow.
We start with flows II and III, because they both work on the off-diagonal alpha tensor with a memory effect,
but one flow has a symmetric and the other an antisymmetric alpha tensor.
We next show the results for flows I and IV, one growing due to the alpha effect (i.e.\ $\mathcal{\boldsymbol{E}}_{\rm emf}\sim \bm{B}$) and the other due to negative diffusivity
(i.e.\ $\mathcal{\boldsymbol{E}}_{\rm emf}\sim +\bm{J}$ on large scales, so instead of a resistive decay, there is growth).

\begin{figure}
    \centering
    \includegraphics[width=0.9\linewidth]{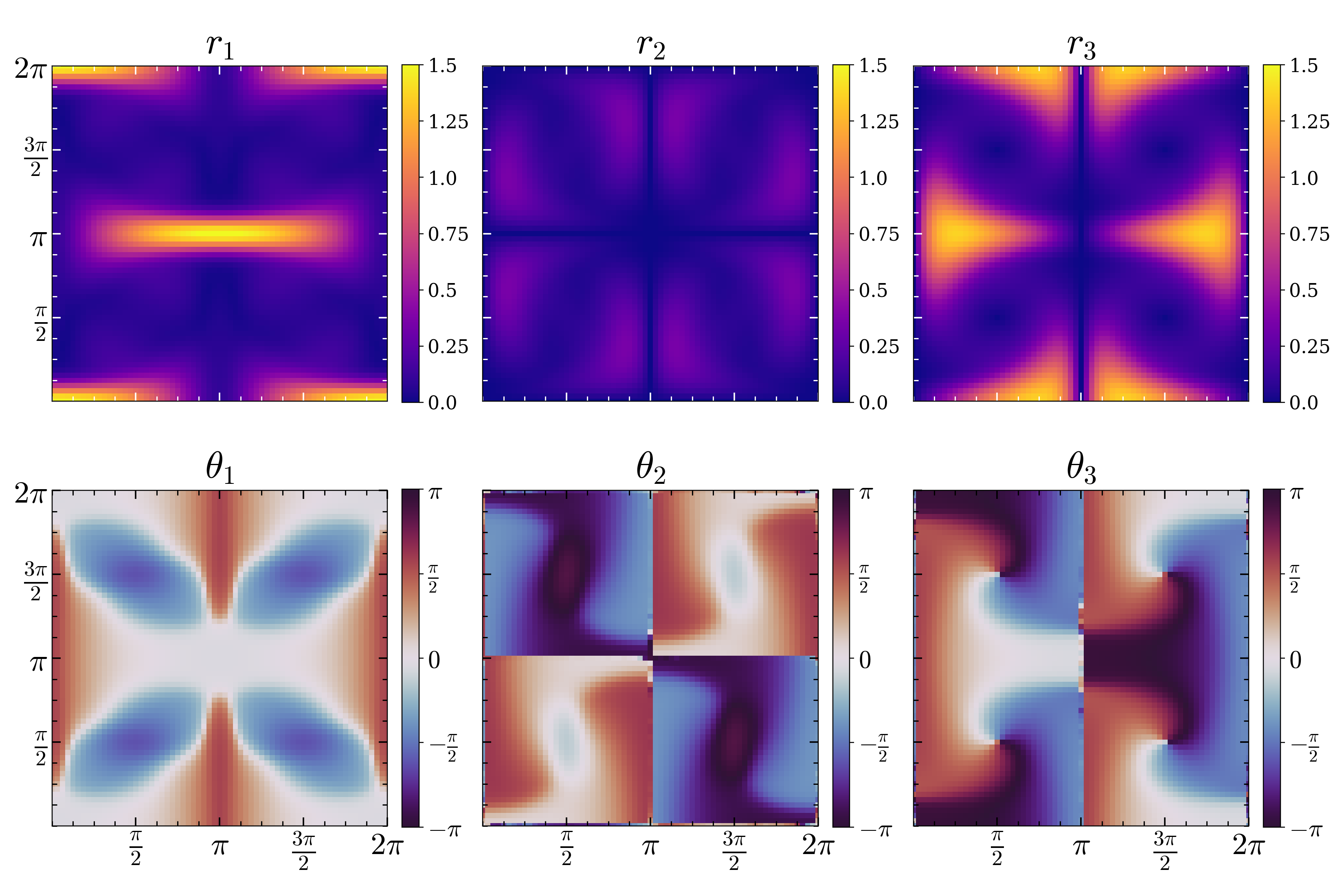}
    \caption{Roberts flow II magnetic field components for the \textsc{Pencil} 2nd-order scheme. The top panel shows $r_1,r_2,r_3$ - maps of the magnitudes for each direction, and the bottom panel shows $\theta_1,\theta_2,\theta_3$ - maps of the complex phase for each component. The phases are fixed by multiplying every $B_i$ by the same phase factor such that $\mathrm{Im} \mean B_x  = 0$.
The modulus-phase decomposition follows \Eq{ModulusPhase}.
The initial conditions follow \Eq{equation:initial_condition} with $k_z=0.5$ giving $\mean B_x \neq0, \mean B_y =0$.
The run was performed with an elongated box ($L_x=L_y=2\pi, L_z=4\pi$) with $64^2\times128$ resolution elements,
with $z$ direction velocity constant $w_0=v_0$ and resistivity of $\eta=0.05$.
The image resolution coincides with the grid resolution in $xy$ plane.}
    \label{fig:pixel_Pencil_rf2dn_064_Run2_c1_BxSin_64_128}
\end{figure}

The results for the Roberts flows II and III runs with $\eta=0.05$, $w_0= v_0$ using the initial condition from \eq{equation:initial_condition}
with $k_z=0.5$ are shown in \Figs{fig:pixel_Pencil_rf2dn_064_Run2_c1_BxSin_64_128}{fig:pixel_Pencil_rf3dn_064_Run2_c1_BxSin_64_128}.
Time independent patterns form in both cases.
The top row of the figures shows density maps in the $xy$ plane for the magnetic field amplitudes of each component; the bottom rows show phases for each magnetic field component,
following the decomposition from \eq{ModulusPhase}.
Roberts flows~I and IV can be initialised with the same initial condition.
For the same $\eta=0.05$, the value of $k_z=1.0$ for flow I and $k_z=0.5$ for flow IV were chosen to excite the fastest growing mode.
The resulting time-independent patterns are illustrated in \Figs{fig:pixel_Pencil_rf1dn_064_Run2_c1_BxSin_64_128}{fig:pixel_Pencil_rf4dn_064_Run2_c1_BxSin_64_128}. 

The $r_1$ map, i.e., the magnitude of the magnetic field in the $x$ direction, shows an elongated bar in the $x$ direction in the shear zone.
This is the stretching region where the magnetic field is amplified.
Because the phase $\theta_1$ stays near zero at corresponding places (and the same for each bar), we can conclude that the magnetic field is pointing in the same direction for each bar.
This means that a non-zero in-plane magnetic field $\tilde B_x$ is present.
This is expected because we explicitly start with an initial condition seeding a non-zero $\tilde B_x$.
We note that for flow~IV the bar is more prominent, but it is likely because for this flow $v_{\rm rms}=1$, unlike the value $v_{\rm rms}=0.866$ for the other flows.

The $r_2$ map, i.e., the magnitude of the magnetic field in the $y$ direction, shows elongated bars only in the case of flow~I.
This is a characteristic feature which, combined with the chosen initial conditions, shows the presence of the alpha effect:
the evolution of magnetic field components is coupled, and starting with $\tilde B_x$ will inevitably create $\tilde B_y$ and the other way around.
In addition, the overall phase is shifted by $\pi/2$ across all planes,
i.e., in 3D there are periodic bars appearing at different heights, and bars along the $x$ and $y$ directions appear one after the other in the $z$ direction.
This is unlike the other flows, where the evolutions of $\tilde B_x$ and $\tilde B_y$ are independent of each other.
Indeed, for flows~II, III, and IV, there are no elongated bars forming.
For flows~II and III, there is a phase change of $\pi$ for neighbouring vortices, such that the magnetic fields contribute with a different sign, resulting in zero total $\tilde B_y$.
Similarly for flow~IV, there is a phase change of $\pi$ in the $y$ direction, making the plane-averaged $ \tilde B_y$ zero.

The $r_3$ map, i.e., the magnitude of the magnetic field in the $z$ direction, is different for each flow.
For example, for flows II and III, the maps of $r_1,r_2,\theta_1,\theta_2$ are mostly similar, and the main difference appears in the $ z$-component geometry.
For flow~I, the $z$-component $r_3$ is much less significant than $r_1$ and $r_2$.
For all other flows, the amplitudes of all three values of $r_i$ are comparable.
For flow~IV, bar-like features appear in the same places as for the $r_1$ component.
However, there is no plane-averaged $z$ field because the bars change phase by $\pi$; that is, they have the opposite magnetic field direction when moving in the $y$ direction.
Flows~II and III do not show bars; both have the features that experience a $\theta_3$ phase change of $\pi$ when moving in the $x$ direction,
i.e., the magnetic field in the $z$ direction changes sign there.
The most significant difference between flows II and III magnetic field patterns comes from the $r_3$ distribution.
While for flow II the magnetic field in the $z$ direction is almost zero inside the vortices, in flow III there is a significant magnetic field in the $z$ direction filling the vortices.
We note also that, for all four flows, each vortex has a continuous circular $\theta_3$ phase shift.

Together, each of these figures shows unique properties of the growing magnetic field spatial distribution for each flow at fixed resistivity
and thus are excellent fingerprints for the effects given in \Tab{Tflows}   

%FIG7
%FIG8
\begin{figure}
    \centering
    \includegraphics[width=0.9\linewidth]{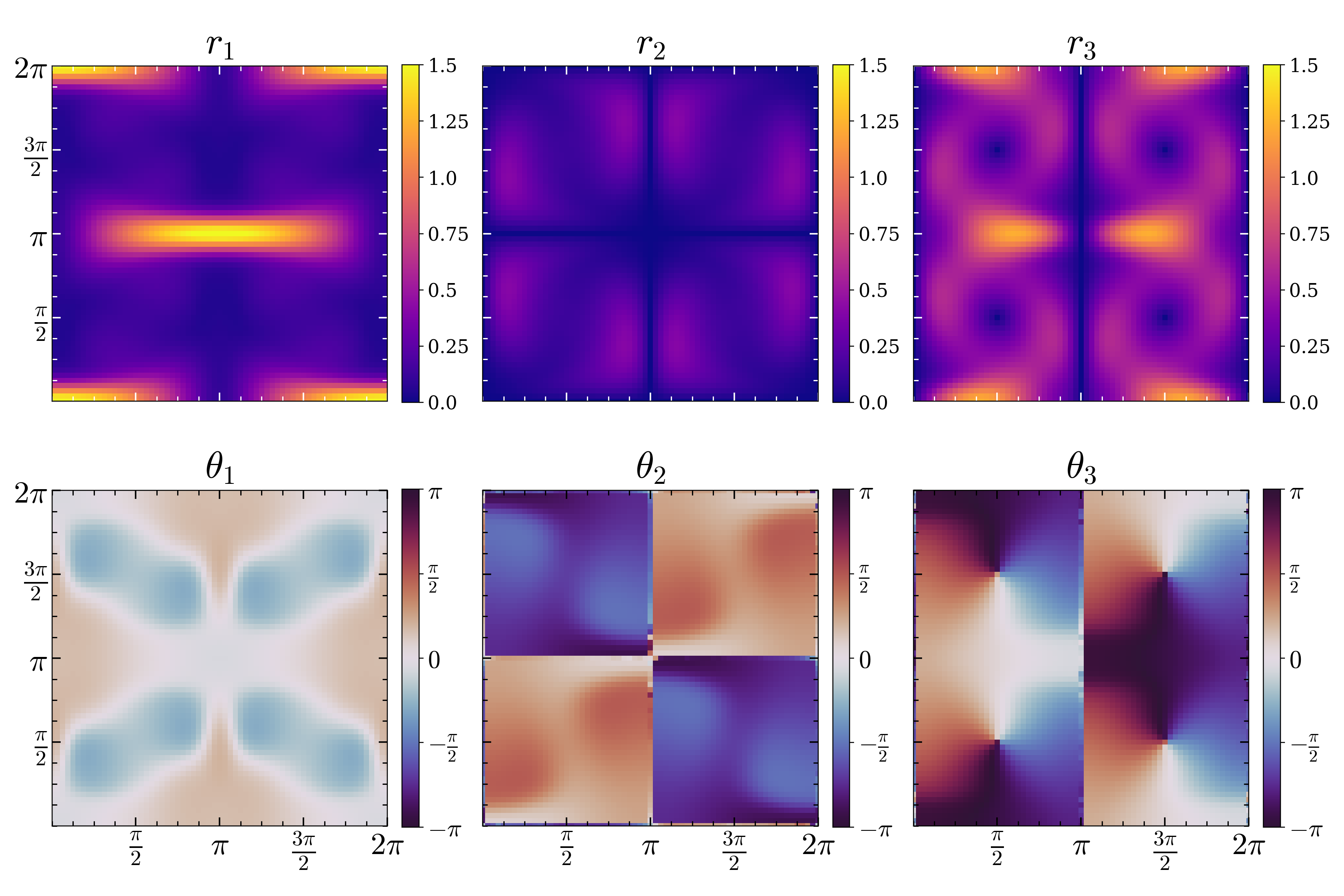}
   \caption{The magnetic field amplitudes and phases for Roberts flow III. The setup is similar to the run from \Fig{fig:pixel_Pencil_rf2dn_064_Run2_c1_BxSin_64_128}}
    \label{fig:pixel_Pencil_rf3dn_064_Run2_c1_BxSin_64_128}
\end{figure}

%FIG9
\begin{figure}
    \centering
    \includegraphics[width=0.9\linewidth]{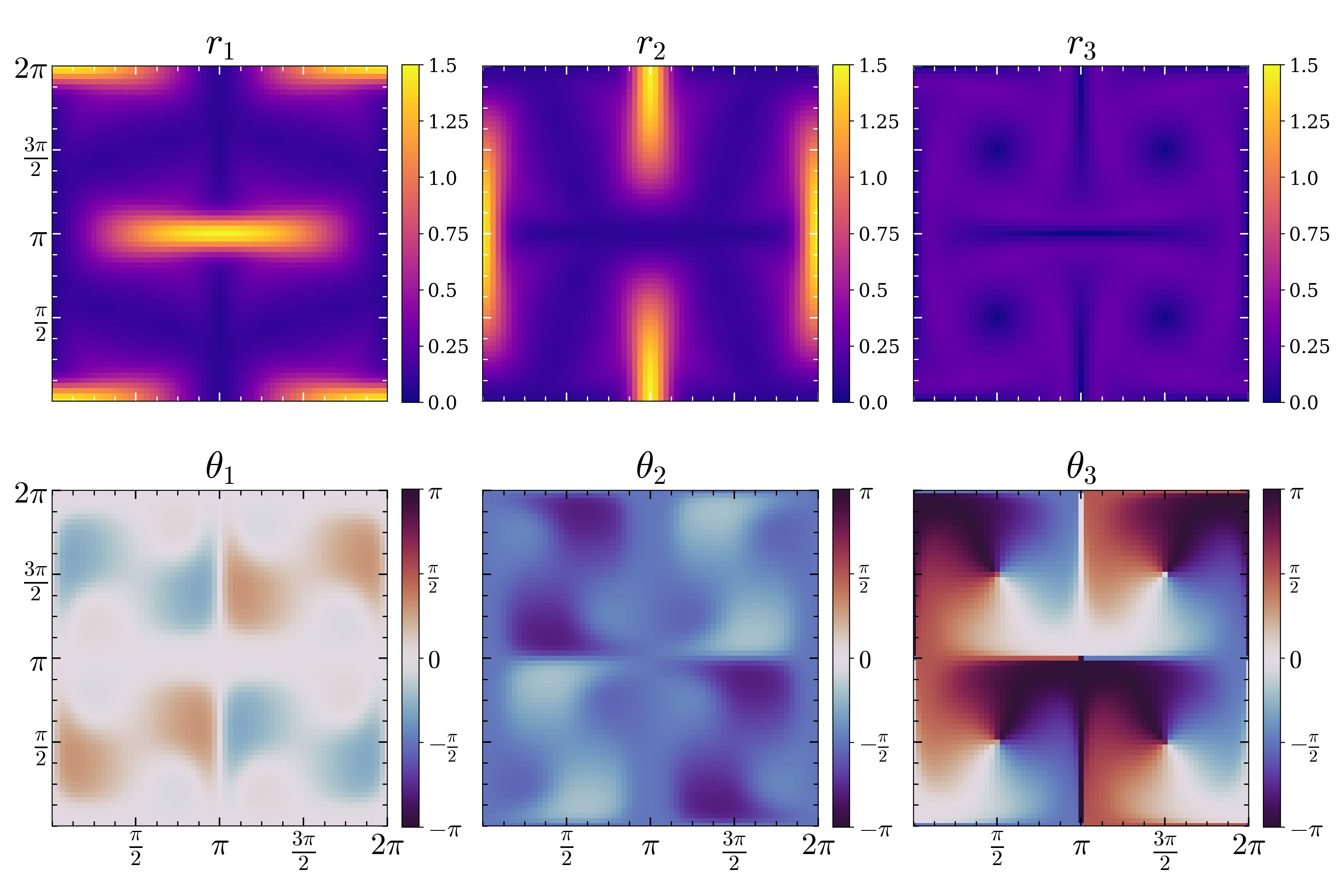}
    \caption{Roberts flow~I magnetic field components for the \textsc{Pencil} 2nd-order scheme.
The top panel shows $r_1,r_2,r_3$ maps of the magnitudes for each direction, and the bottom panel shows $\theta_1,\theta_2,\theta_3$ maps of the complex phase for each component.
The phases are fixed by multiplying every $B_i$ by the same phase factor such that $\mathrm{Im} \mean B_x  = 0$.
The modulus-phase decomposition follows \Eq{ModulusPhase}.
The initial conditions follow \Eq{equation:initial_condition} with $k_z=1.0$ giving $\mean B_x \neq0, \mean B_y =0$. The run was performed with the elongated box ($L_x=L_y=2\pi, L_z=4\pi$) with $64^2\times128$ resolution elements, with $ z$-direction velocity constant $w_0=v_0$ and resistivity of $\eta=0.05$. }
    \label{fig:pixel_Pencil_rf1dn_064_Run2_c1_BxSin_64_128}
\end{figure}

%FIG10
\begin{figure}
    \centering
    \includegraphics[width=0.9\linewidth]{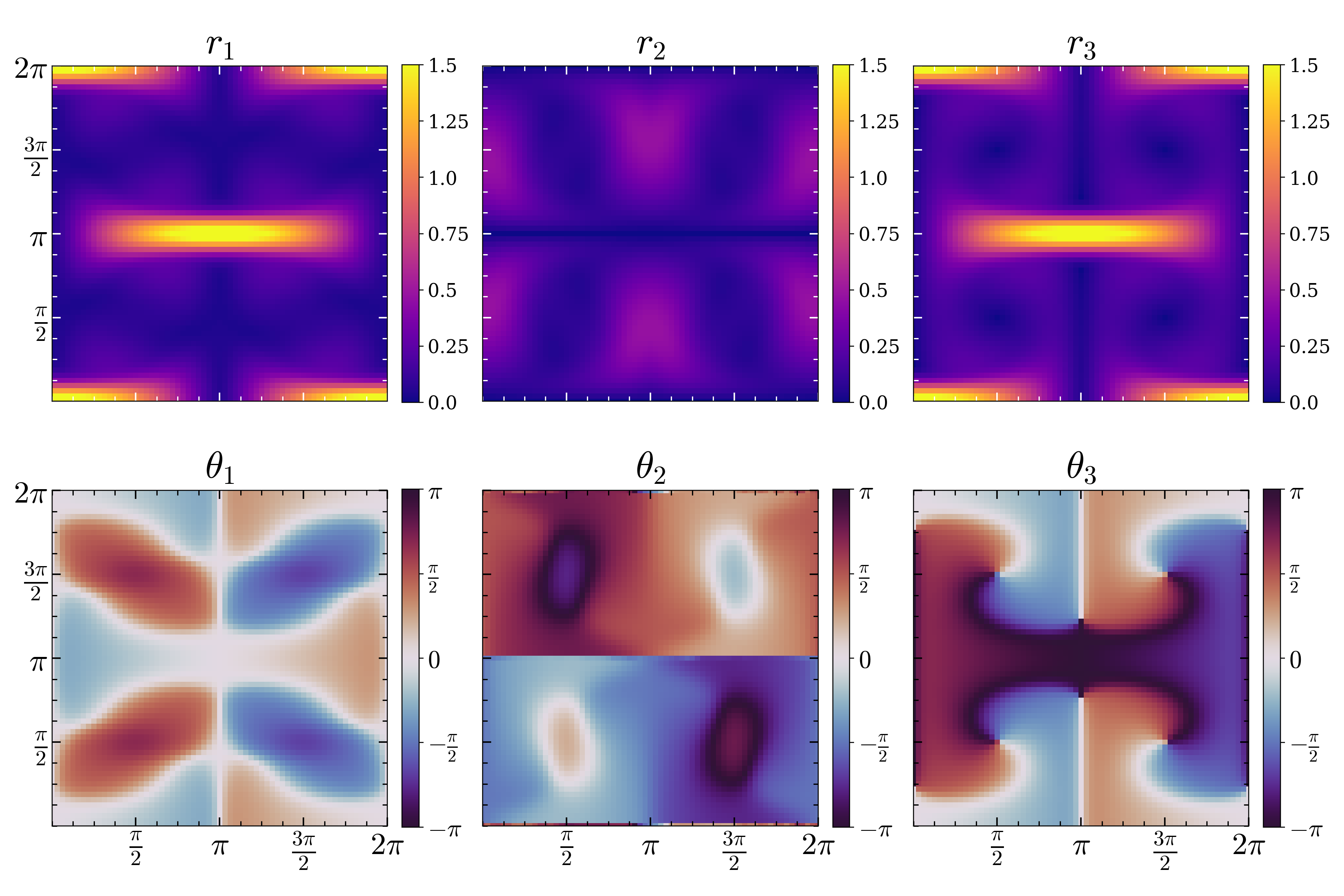}
    \caption{The magnetic field amplitudes and phases for Roberts flow IV. The setup is similar to the run from \Fig{fig:pixel_Pencil_rf1dn_064_Run2_c1_BxSin_64_128} with initial $k_z=0.5$}    \label{fig:pixel_Pencil_rf4dn_064_Run2_c1_BxSin_64_128} 
\end{figure}

\subsection{Convergence of the magnetic field patterns} \label{sec:convergence_patterns}
\begin{figure}
    \centering
    \includegraphics[width=0.95\linewidth]{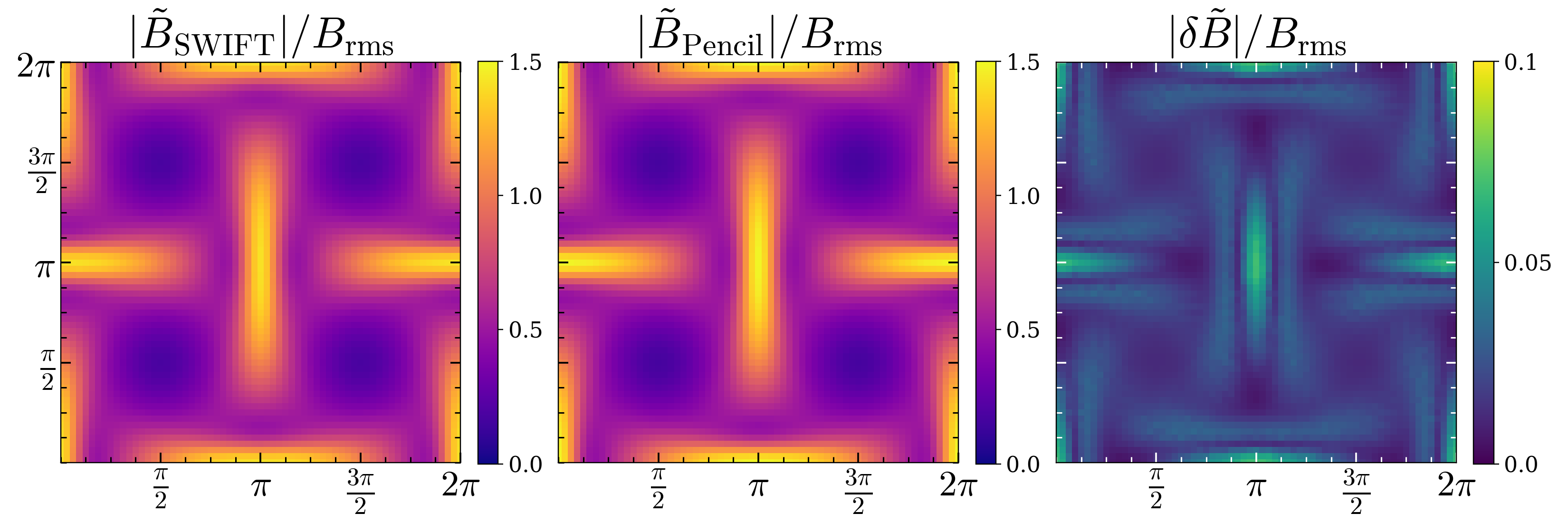}
    \caption{Maps for absolute values of the complex magnetic field (eq. \ref{PatternAbs}) for \textsc{swift}, \textsc{pencil} and a map of the absolute value of the difference between two complex fields (eq. \ref{PatternDifference}). The runs were performed with $w_0=v_0$. The error is largest at the peaks of the features and at the sides -  consistent with an assumption that \textsc{swift} produces smoother features at the same number of resolution elements.}
    \label{fig:SWIFT_r1_comparisons}
\end{figure}

As scalar quantities of interest, the magnetic field patterns should converge in the continuum limit.
Therefore, the difference in the magnetic field patterns for both codes should vanish with resolution.
In what follows below, we compare the magnetic field patterns for the Roberts flow~I.
We simplify the comparison by choosing the absolute value of the magnetic fields rather than using the components separately.
We define the scalar difference between magnetic field patterns using \Eq{PatternDifference} and study the convergence of the difference with numerical resolution.
Figure~\ref{fig:SWIFT_r1_comparisons} shows the absolute values of the complex maps [see \Eq{PatternAbs}]
for \textsc{swift} (first panel) and \textsc{pencil} (second panel).
The patterns are almost visually indistinguishable.
The difference between the two magnetic field distributions, quantified through \eq{PatternDifference},
is depicted in the third panel of \Fig{fig:SWIFT_r1_comparisons}.
The difference is at the $\sim 5\%$ level and concentrated near the centres and the edges of the features.
This is consistent with an assumption that features in \textsc{SWIFT} are more smooth than the corresponding features
in the \textsc{Pencil} second-order scheme for a fixed number of resolution elements.

\begin{figure}
    \centering
    \includegraphics[width=0.5\linewidth]{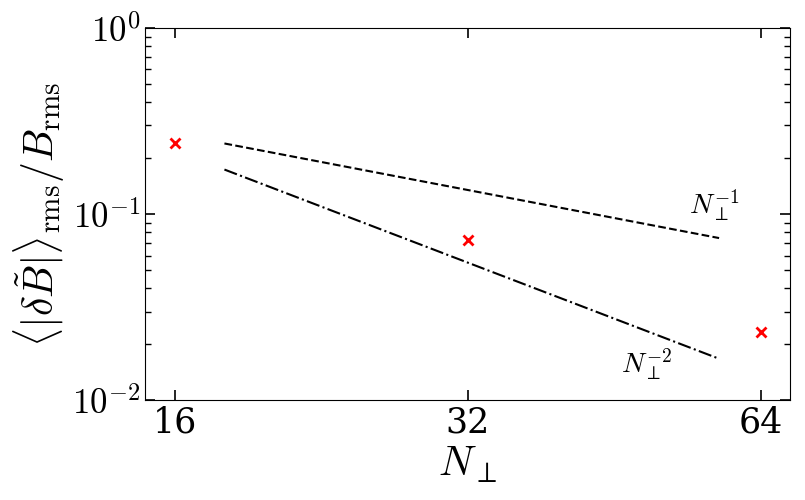}
    
    \caption{Convergence of the difference norm \eq{PatternDifferenceRMS} of the magnetic field pattern for Run~2 for \textsc{SWIFT} with respect to the same-resolution \textsc{Pencil} 2nd-order runs.
The runs were performed at $\eta=0.05$ for Roberts flow~1 with $w_0=v_0$.
The norms for each magnetic field component is denoted with plusses, crosses, and dots.
Convergence is mostly second order in the number of resolution elements.}
    \label{fig:L2_swift_pencil_B}
\end{figure}

We use \Eq{PatternDifferenceRMS} to quantify the plane pattern difference with a single number for each resolution.
Figure~\ref{fig:L2_swift_pencil_B} shows the magnitude of the pattern error as a function of resolution.
The lines indicate the $N_\perp^{-2}$ and $N_\perp^{-1}$  slopes.
The pattern difference converges with resolution roughly as $N_{\perp}^{-2}$, although somewhat more slowly.
One possibility for this deviation could be that the comparison analysis consists of many steps,
such as transforming \textsc{swift} particle data onto a grid, which is then compared with the \textsc{Pencil} result.
Another possibility is the presence of $N_\perp^{-1}$ contributions in the particle distribution in the \textsc{swift} case.
The differences for the growing solutions in the 3D magnetic field distributions between the \textsc{Pencil} 2nd order scheme and the \textsc{SWIFT} code vanish as $N_\perp^{-2}$.
This conclusion will likely stay the same for runs with larger and smaller values of resistivity and for any other flow,
if resistivity is large enough to prevent the resolution-related issues mentioned in \Sec{sec:spacial_resolution_limits},
because the difference between the codes reduces to a smoother magnetic field at the resolution scale; therefore, the difference decreases at higher resolutions.

%FIG12

\section{Conclusions}

%\section*{Acknowledgements}

The four Roberts flows represent distinct, simple 2D dynamo-capable flows illustrating different mechanisms for large-scale slow dynamos.
The first flow demonstrates the alpha tensor, the second and third flows work on the off-diagonal alpha effect with memory effect,
and the fourth flow does not exhibit the alpha effect and grows due to a negative turbulent diffusivity.
The system is well studied semi-analytically and with direct numerical simulations.
We present the results for the \textsc{Pencil} and \textsc{SWIFT} codes with parameters, where a comparison for all four flows is possible.
The codes represent extremely different approaches to discretising MHD equations:
one solves the equations on a fixed grid with vector potentials, and the other solves in the frame of moving particles by evolving the $\bm{B}$ field directly.
Yet, in the region where they are applicable, both \textsc{Pencil} and \textsc{SWIFT} recover the expected features.
Our results can be summarised as follows.
\begin{itemize}
    \item We have presented the results for the growth rate resistivity study in \Fig{fig:roberts_flows_growth_rates}.
A very broad set of features in Roberts flows is recovered in both \textsc{Pencil} and \textsc{swift}, such as the critical magnetic Reynolds numbers, the mode transitions,
the low-resistivity behaviour for flow~I and the oscillatory solutions for flow~IV.
We note that \textsc{swift} produced larger growth rates at the same number of resolution elements (particles or grid points) and resistivity.
Consequently, the critical magnetic Reynolds numbers shift to lower values.
    \item To quantitatively describe the growing solutions,
we have employed the growth rate, the plane-averaged magnetic field, the magnetic Taylor microscale,
the work against the Lorentz force, and the various products of $\bm{J}$ and $\bm{B}$.
The resulting dependences of the quantities on resistivity and the flow are shown in \Tabs{Tresults}{Tresults2}.
We provide the reference values for the high-resolution ($256^2\times 512$) runs for  \textsc{Pencil} and \textsc{SWIFT} in \Tab{TresultsPencilSwift}.
    \item We have studied the convergence for a fixed value of the resistivity $ \eta = 0.05$.
Figure \ref{p5em2} shows that the quantities of interest converge with resolution.
For \textsc{Pencil}, the quantities converge roughly as the order of the corresponding scheme: 10th order converges as $N_\perp^{-10}$, 6th order as  $N_\perp^{-6}$, 2nd order as  $N_\perp^{-2}$.
\textsc{SWIFT} code convergence has the same slope as the second-order \textsc{Pencil} line, but is shifted towards higher resolution. To achieve error levels comparable to those of the second-order \textsc{Pencil} scheme, \textsc{SWIFT} requires three times as many resolution elements per dimension.

We note that \textsc{SWIFT} converges to values different from \textsc{Pencil}.
We think one contribution to this discrepancy could be a suboptimal particle distribution and density fluctuations which appear in shear regions,
where the magnetic field grows, which can cause errors in the derivative operators in the induction equation.
    \item We have only studied the spatial resolution convergence for both \textsc{Pencil} and \textsc{SWIFT}. However, the time resolution convergence and sound speed dependence can also impact the results.
    \item We have found that a convenient choice of initial conditions is one where the maps of the complex magnetic field can be built to display a unique,
time-independent magnetic field patterns describing the 3D field distribution for all four flows.
The patterns are shown in \Figss{fig:pixel_Pencil_rf2dn_064_Run2_c1_BxSin_64_128}{fig:pixel_Pencil_rf4dn_064_Run2_c1_BxSin_64_128}.
Such a representation allows for a per-pixel comparison between magnetic field configurations obtained with different codes.
    \item The patterns can be compared based on the absolute values of the complex magnetic field.
The main difference between the 2nd-order \textsc{Pencil} and \textsc{SWIFT} results at the same number of resolution elements
is attributed to the magnetic field being smoother for \textsc{SWIFT} code, as illustrated on \Fig{fig:SWIFT_r1_comparisons}.
We used the difference metric [see \Eq{PatternDifferenceRMS}] to study how the pattern difference between \textsc{swift} and \textsc{pencil}
changes with linear resolution, which turned out to converge as $N_\perp^{-2}$, as shown in \Fig{fig:L2_swift_pencil_B}.
    \item The conclusions about the code convergence and pattern difference will very likely stay similar for any other chosen resistivity (but large enough to prevent overwinding) and for any of the four flows.
    \item The features in the growth rate versus resistivity dependence from \Fig{fig:roberts_flows_growth_rates}, the reference values from \Tabs{Tresults}{Tresults2},
and the resolution convergence of the quantities from \Sec{sec:observable_characteristic_quantities}, as shown in \Fig{p5em2},
together with the unique fingerprint patterns characterising the growing magnetic field from \Figss{fig:pixel_Pencil_rf2dn_064_Run2_c1_BxSin_64_128}{fig:pixel_Pencil_rf4dn_064_Run2_c1_BxSin_64_128} for all four Roberts flows can be used as a benchmark for the future code comparisons in their ability to reproduce features of large-scale kinematic dynamos.
\item We also confirm that resistivity cannot be chosen arbitrarily low for both \textsc{Pencil} and \textsc{SWIFT} codes.
While flow II has the smallest growth rate, it is the most sensitive one to crossing the conservative limit on resistivity.
Outside this limit, both codes produce spurious growth, while employing higher resolution solves the issue.
This highlights the importance of controlling the smoothness of magnetic field features at the resolution scale, by, for example, adding enough physical resistivity.
 \item We also note that in \textsc{Pencil}, the difference in the computational time between sixth-order and tenth
order schemes is relatively minor (about 40\%).
Thus, it is generally advantageous to use higher-order representations.
It is often thought that higher-order schemes are disadvantageous when
shocks form.
This is, however, not the case when shocks are considered as spatially
resolved, sharp structures.
This was already demonstrated in earlier work \citep{Bra03}.
In fact, as long as the structures are spatially smooth, they can more
accurately be represented with fewer points in a scheme of higher order.
This again highlights the importance of high-order schemes.
\end{itemize}

\section*{Funding}
%This work was supported in part by the Swedish Research Council (Vetenskapsr{\aa}det, 2025-05957),
This work was supported in part by the European Research Council through the ERC Synergy Grant COSMOMAG under grant No.\ 101224803,
the Swedish Research Council (Vetenskapsradet, 2025-05957),
the National Science Foundation under grant no.\ NSF AST-2307698 and a NASA ATP Award 80NSSC22K0825.
We acknowledge the inspiring atmosphere during the program on the
``Generation, evolution, and observations of cosmological magnetic fields''
at the Bernoulli Center in Lausanne.
We acknowledge the allocation of computing resources provided by the
Swedish National Allocations Committee at the Center for Parallel
Computers at the Royal Institute of Technology in Stockholm and
Link\"oping.

\section*{Declaration of Interests}
The authors report no conflict of interest.

\section*{Data availability statement}
% The data that support the findings of this study are openly available on
% %Zenodo at doi:10.5281/zenodo.10527437 (v2024.11.02) or, for easier access, at
% \url{http://norlx65.nordita.org/~brandenb/projects/Convergence-Roberts}.\\
% All calculations have been performed with the {\sc \textsc{Pencil} Code}
% \citep{JOSS}; \url{https://github.com/pencil-code}; DOI:10.5281/zenodo.3961647,
%  and {\sc \textsc{SWIFT}} code \citep{SWIFT,Karapiperis_2026} - entirely public, including the examples presented in
% this work. It can be found alongside an extensive documentation on the website
% of the project: \url{www.swiftsim.com}.

The data supporting the findings of this are openly available on \url{http://norlx65.nordita.org/~brandenb/projects/Convergence-Roberts}.\\

All calculations were performed using {\sc Pencil Code} \citep{JOSS} and {\sc SWIFT} code \citep{SWIFT,Karapiperis_2026}. Both codes are publicly available, including the examples used in this work. The {\sc Pencil Code} is available at \url{https://github.com/pencil-code} (DOI: 10.5281/zenodo.3961647). The {\sc SWIFT} code, together with extensive documentation and example data, is available at \url{https://www.swiftsim.com}.

\section*{Authors' ORCIDs}

\noindent
N. Shchutskyi https://orcid.org/0009-0005-5547-7580\\
A. Brandenburg, https://orcid.org/0000-0002-7304-021X\\
M. Schaller https://orcid.org/0000-0002-2395-4902

\bibliography{ref}{}
\bibliographystyle{jpp}

\appendix
\section{Time integration setup for Roberts flows in \textsc{SWIFT}}
Regarding the time resolution, in the \textsc{SWIFT} code, there are two time step conditions relevant to the Roberts flow runs.
The Courant-Friedrich-Levy (CFL) condition: 
\begin{equation}
    dt_\mathrm{CFL} = C_\mathrm{CFL} \, h / v_\mathrm{sig},
\end{equation}
which, in the case of an isothermal equation of state, depends on the fixed sound speed ($v_\mathrm{sig}=c_\mathrm{s}$) and is therefore set manually ($c_s = 1$, $C_\mathrm{CFL} \simeq 0.075$),
and the resistive time step:
\begin{equation}
    dt_\eta = C_\eta \, h^2/\eta,
\end{equation}
where for the $64^3$ run we use $C_{\eta} \simeq \frac{1}{8\pi}$.
We employ a quintic spline kernel with 113 neighbours (such that $h = 1.36 \, (m/\rho)^{1/3}$).
For a box with dimensions $L_\perp^2 \times L_\|$, and $L_\perp = 2 \pi$, the smoothing length is $h = 1.36 \times 2\pi/64\simeq 0.13$.
This results in the ratio:
\begin{equation}
    \frac{dt_\mathrm{CFL}}{dt_\eta}\simeq \frac{\eta}{0.15}.
\end{equation}
Therefore, for resistivities larger than $\eta \gtrsim 0.15$, the resistive time step constraint matters.

For the runs above, we chose $c_\mathrm{s} = 1$, which provides a significant speed-up compared to the choice $c_s=33$ used in \cite{SSKSB25}.
However, such a choice does impact the growth rate accuracy and generally produces a few per cent larger growth rates
in \Tab{TresultsPencilSwift}.
Another hint pointing at the time step importance is the noticeable decrease in the deviation from the \textsc{Pencil} growth rate result
from \Fig{fig:roberts_flows_growth_rates} for flow~IV near $\eta \simeq 0.1$, which is roughly where the transition to resistive time step happens.

\label{appx:SWIFTdt}

\section{Fit parameters for the convergence plots}
\label{appx:convergence_fit}
We performed a linear fit for the high-resolution part of the convergence study, shown in \Fig{p5em2}.
We used the following parametric form for the quantity error, defined in \eq{eq:error_def}:
\begin{equation}
 \mathrm{err}(X,N_\perp) = (N_\perp / N_X)^{-n_X}.
 \label{eq:err_fit}
\end{equation}
We fit the last two points ($N_\perp = 64,128$) for the 6th and 10th order \textsc{Pencil} code,
and the last 3 points ($N_\perp =32,64,128$) for the \textsc{SWIFT} and 2nd order \textsc{Pencil} schemes.
The results are provided in \Tab{TConvFit}.
In addition, from the convergence of the 10th order scheme one can deduce the required accuracy for the reference run.
Among all displayed quantities, we find that the growth rate for the reference 10th-order \textsc{Pencil} run
is, owing to the high accuracy, the most demanding quantity with the error being (deduced from the fit):
\begin{equation}
    \mathrm{err}(\lambda,N_\perp=128) \sim  10^{-11}, \qquad \mathrm{err}(\lambda,N_\perp=256) \sim  10^{-14}.
\end{equation}
Therefore, because the error for the $N_\perp = 128$ run is in the 11th digit, in order to reconstruct the convergence shown in \Fig{p5em2},
the accuracy of the calculated growth rate for the reference run at $N_\perp = 256$ must be much better. Using the same scaling, one can expect the error to be in the 14th digit. However, conservatively, it is enough to have the result accurate up to 12 digits. For this reason, the results for the reference run are shown in \Tab{TresultsPencilSwift} up to 12 digits.
\begin{table}\caption{
Results for the convergence fit using \Eq{eq:err_fit}.
}\vspace{12pt}\centerline{\begin{tabular}{llrr}
Scheme & Quantity & $n_X$ &  $N_X$ \\
\hline
Pencil 10th & $\lambda/v_0 k_0                        $ & 10.0 & 10.4 \\
 & $\meanBrms/\Brms                        $ & 9.6 & 11.6 \\
  & $-W_\mathrm{L}/\urms\Brms^2         $ & 9.6 & 15.8 \\
  & $\Jrms/\Brms $ & 9.6 & 18.1 \\
  &$\bra{(\JJ\cdot\BB)^2}^{1/2}/\Brms^2$ & 9.8 & 15.3 \\
  &$\bra{\JJ\cdot\BB}/\Brms^2$ & 10.0 & 11.5 \\
  \hline
Pencil 6th & $\lambda/v_0 k_0                        $ & 5.9 & 6.6 \\
 & $\meanBrms/\Brms                        $ & 5.9 & 7.4 \\
  & $-W_\mathrm{L}/\urms\Brms^2         $ & 5.8 & 11.8 \\
  & $\Jrms/\Brms $ & 5.8 & 14.5 \\
  &$\bra{(\JJ\cdot\BB)^2}^{1/2}/\Brms^2$ & 5.9 & 11.7 \\
  &$\bra{\JJ\cdot\BB}/\Brms^2$ & 5.9 & 7.8 \\
\hline
Pencil 2nd & $\lambda/v_0 k_0                        $ & 2.1 & 1.5 \\
 & $\meanBrms/\Brms                        $ & 1.8 & 0.6 \\
  & $-W_\mathrm{L}/\urms\Brms^2         $ & 2.0 & 5.6 \\
  & $\Jrms/\Brms $ & 2.0 & 6.7 \\
  &$\bra{(\JJ\cdot\BB)^2}^{1/2}/\Brms^2$ & 2.4 & 5.0 \\
  &$\bra{\JJ\cdot\BB}/\Brms^2$ & 2.2 & 3.6 \\
\hline
SWIFT & $\lambda/v_0 k_0                        $ & 2.0 & 4.8 \\
 & $\meanBrms/\Brms                        $ & 1.9 & 3.7 \\
  & $-W_\mathrm{L}/\urms\Brms^2         $ & 0.9 & 0.1 \\
  & $\Jrms/\Brms $ & 1.0 & 3.8 \\
  &$\bra{(\JJ\cdot\BB)^2}^{1/2}/\Brms^2$ & 1.8 & 13.3 \\
  &$\bra{\JJ\cdot\BB}/\Brms^2$ & 1.9 & 18.9 \\
  
\label{TConvFit}\end{tabular}}\end{table}

\section{Importance of $N_\|$ for \textsc{Pencil} Code runs}

\label{appx:Npar}

\begin{figure}
\begin{center}
\includegraphics[width=0.9\columnwidth]{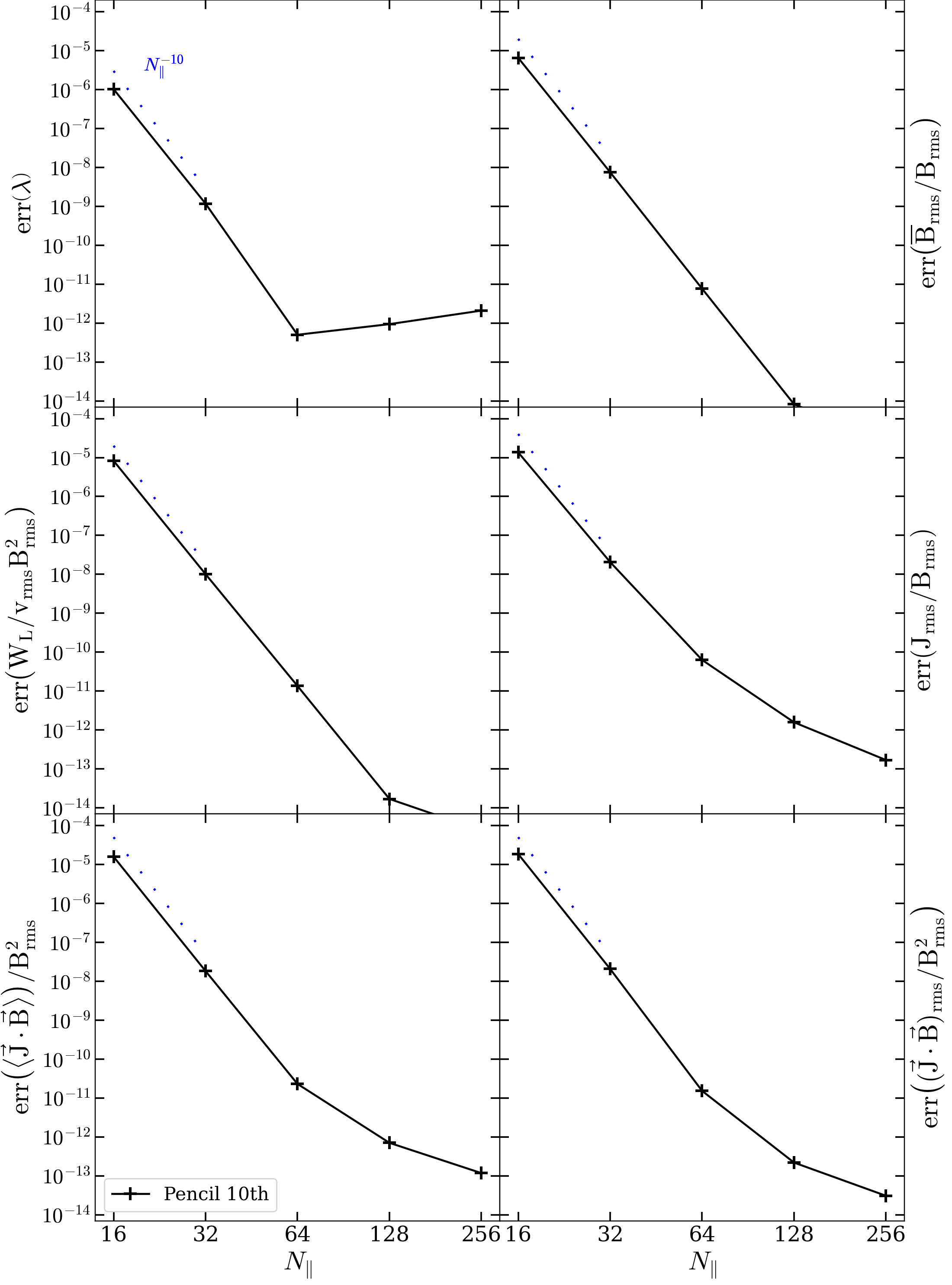}
\end{center}
\caption[]{
Convergence properties of different quantities for
tenth-order \textsc{Pencil} scheme with the number of resolution elements in the z-direction, $N_\|$. Convergence is shown for \textsc{Pencil} with respect to its highest resolution $256^2\times 512$ 10th-order result. The longer dotted line denotes $1/N_\|^{10}$ slope.
}\label{p5em2_Npar}\end{figure}

The \textsc{Pencil Code} convergence with resolution was studied by employing uniform sampling.
However, because of the chosen value of the resistivity, the dependence on the $z$ direction is a harmonic wave solution with $k_z = 1.0$,
the features in the $z$ direction will be much smoother than the dependence on the $x$ and $y$ directions,
and it is not necessary to choose $N_\| = 2 N_\perp$ when $L_\|=2L_\perp$.
In \Fig{p5em2_Npar} the convergence with $N_\|\in [16,256]$ with fixed $N_\perp = 256$ relative to the reference $256^2\times 512$ run is shown.
All quantities initially converge as $N_\|^{-10}$.
The growth rate stops improving for the choice $N_\|=64$ and $N_\perp=256$.
The errors for other quantities change the slope but still continue improving.
For smaller values of the resistivity, $k_z$ can increase, and more uniform sampling may be required.

%FIG5

\end{document}